\documentclass[twocolumn,showpacs,nofootinbib,preprintnumbers,prd]{revtex4-1}

\usepackage{amsmath}
\usepackage{amsfonts}
\usepackage{amssymb}
\usepackage{graphicx}
\usepackage[titletoc]{appendix}
\usepackage{color}
\usepackage{hyperref}
\usepackage{cleveref}
\usepackage[rightcaption]{sidecap}
\usepackage{subcaption}
\usepackage{comment}

\usepackage{mathtools}

\usepackage{dcolumn}

\usepackage{array}
\usepackage{ctable}
\usepackage{multirow}
\usepackage{siunitx}
\usepackage{tabularx}
\usepackage{booktabs}
\usepackage{supertabular}
\usepackage{verbatim}

\usepackage{cancel}
\usepackage{graphicx}
\usepackage{dcolumn}
\usepackage{bm}
\usepackage{braket}

\begin{document}

\title{Impact of a complex scalar spectator field on baryon asymmetry within spontaneous baryogenesis}

\author{Mattia Dubbini}
\email{mattia.dubbini@studenti.unicam.it}
\affiliation{Universit\`a di Camerino, Via Madonna delle Carceri, Camerino, 62032, Italy.}

\author{Orlando Luongo}
\email{orlando.luongo@unicam.it}
\affiliation{Universit\`a di Camerino, Via Madonna delle Carceri, Camerino, 62032, Italy.}
\affiliation{Department of Nanoscale Science and Engineering, University at Albany-SUNY, Albany, New York 12222, USA.}
\affiliation{Istituto Nazionale di Astrofisica (INAF), Osservatorio Astronomico di Brera, 20121 Milano, Italy.}
\affiliation{Istituto Nazionale di Fisica Nucleare (INFN), Sezione di Perugia, Perugia, 06123, Italy,}
\affiliation{Al-Farabi Kazakh National University, Al-Farabi av. 71, 050040 Almaty, Kazakhstan.}

\author{Marco Muccino}
\email{marco.muccino@unicam.it}
\affiliation{Universit\`a di Camerino, Via Madonna delle Carceri, Camerino, 62032, Italy.}
\affiliation{Al-Farabi Kazakh National University, Al-Farabi av. 71, 050040 Almaty, Kazakhstan.}
\affiliation{ICRANet, Piazza della Repubblica 10, 65122 Pescara, Italy.}

\date{\today}

\begin{abstract}
We extend the framework of spontaneous baryogenesis by investigating the generation of baryon asymmetry when the inflaton, $\theta$, is minimally coupled with a complex spectator scalar field $\phi$, as $\theta^2|\phi|^2$. To do so, we also consider $\phi$ non-minimally coupled with the Ricci scalar curvature $R$ through a Yukawa-like interaction. We do not consider further interactions of the spectator field with the fermions of the Standard Model, considering it \emph{de facto} as a dark scalar field. In evaluating the violation of the baryon-number conservation during the reheating epoch, in a perfectly homogeneous and isotropic universe, we follow a semiclassical approach, where $\theta$, $\phi$ and gravity are considered as classical fields, whereas the fermions are quantized. We solve the equations of motion for the inflaton and spectator fields, respectively at first and zero-order in perturbation theory, neglecting at first stage the expansion of the universe. Afterwards, we quantify how the spectator field modifies the inflationary dynamics and thus find the baryon asymmetry produced via the inflaton decays into fermion-antifermion pairs by computing the corresponding decay amplitudes. We therefore obtain small first order correction to standard spontaneous baryogenesis and finally discuss the mass-mixing between fermions. Accordingly, the effects of considering the universe expansion are accounted, showing when the coupling between $\phi$ and $R$ becomes noticeable in altering the overall baryon asymmetry.
\end{abstract}

\maketitle
\tableofcontents

\section{Introduction}\label{sec:level1}

The standard Big Bang paradigm and the most accepted background, namely the $\Lambda$CDM model, appear unable to account for the observed baryon–antibaryon asymmetry \cite{Riotto:1998bt} that, once normalized to the universe entropy density, yields $\sim 10^{-10}$--$10^{-11}$.

Theoretical mechanisms invoked to produce the observed baryonic asymmetry fall into the so called baryogenesis theories \cite{Cline:2006ts, Balazs:2014eba, Dimopoulos:1978kv, Dimopoulos:1978pw, Weinberg:1979bt}. Many models \cite{Riotto:1998bt, Cohen:1990it, Farrar:1993hn} have been proposed since the pioneering Sakharov criteria were established \cite{Sakharov:1967dj}.
Among them, we underline Affleck–Dine baryogenesis \cite{Cline:2019fxx, Lloyd-Stubbs:2020sed, Bettoni:2018utf, Bettoni:2018pbl, Bettoni:2019dcw, Bettoni:2021zhq, Laverda:2023uqv, Bettoni:2024ixe}, providing baryogenesis from the out-of-equilibrium evolution of a complex scalar field with a simple renormalizable potential, and baryogenesis via leptogenesis.

In particular, leptogenesis currently furnishes one of the most persuasive and widely supported frameworks \cite{Riotto:1998bt,Fong:2012buy,Davidson:2008bu,Bodeker:2020ghk} in which a lepton asymmetry is generated in the early universe through the CP-violating decays of heavy Majorana neutrinos. The model adopts the \emph{seesaw mechanism} for neutrino mass generation \cite{Dey:2021ecr,Qi:2024pqe,Okada:2012fs,Garny:2006xn} and, consequently, induces -- via lepton asymmetry -- a baryon asymmetry through nonperturbative electroweak processes\footnote{Those are known as sphalerons, violating baryon plus lepton number, $B+L$, while conserving $B-L$.}\cite{Morrissey:2012db}. Leptogenesis may thus provide a natural link between the observed matter–antimatter asymmetry and the neutrino mass origin  \cite{Buchmuller:2003gz}.

However, baryogenesis can be constructed in several other ways \cite{Arbuzova:2018hcj}. In particular, two of the most consolidated models are the \emph{spontaneous baryogenesis} \cite{Dolgov:1994zq,Dolgov:1996qq,Luongo:2021gho} and the \emph{gravitational baryogenesis} \cite{Davoudiasl:2004gf,Arbuzova:2023rri,Sadjadi:2007dx}.

Within the spontaneous baryogenesis framework, in the primordial universe, above a certain energy scale the baryon-number conservation is guaranteed by a global $U(1)$-symmetry, whereas below it the symmetry is spontaneously broken and thus baryon-number violating interactions occur.
In this scenario, the inflaton plays the role of the \emph{pseudo Nambu-Goldstone boson}, whose baryon-number violating decays during reheating into fermion-antifermion pairs \cite{Racker:2014yfa, DeSimone:2016ofp} can produce the experimentally observed baryon asymmetry.
In this framework, recently the effects in terms of baryon asymmetry production of non-minimally coupling the inflaton to the Ricci scalar curvature have been investigated, obtaining rather large corrections to the background model \cite{Dubbini:2025jjz}.

In the more recent scenario of the gravitational baryogenesis \cite{Goodarzi:2023ltp,Mojahed:2024mvb}, baryon asymmetry arises from a coupling between the derivative of the Ricci scalar and a matter current (scalar or fermionic) in thermal equilibrium. This mechanism, however, suffers from instabilities in the Ricci curvature when using the Einstein-Hilbert action, leading to exponentially divergent solutions, suggesting possible stabilizing additional terms, such as $R^2$ \cite{Arbuzova:2017zby}.

In this work, we focus on the promising approach of spontaneous baryogenesis. Hence, in addition to the widely-considered background model \cite{Dolgov:1994zq,Dolgov:1996qq}, we introduce a subdominant complex scalar spectator field, $\phi$, assumed to be less massive and weakly interacting with respect to the inflaton field, $\theta$. We define a possible couple via  $\xi\theta^2|\phi|^2$, where $\xi$ represents the underlying coupling constant. This assumption translates the need of including additional fields into the spontaneous baryogenesis background, since it is quite likely to expect more than one scalar field, namely $\theta$, within the primordial times. Accordingly, the spectator introduction implies a modification of inflaton's mass term within the equations of motion and, correspondingly, it influences the probability of the inflaton decays into fermion-antifermion pairs and lead to a correction to the baryon asymmetry. By construction, such a correction appears small with respect to the background, since the subdominant and lighter spectator field is weakly interacting with the inflaton itself.
To check how the kind of interaction may influence the overall particle production, we even consider a non-minimal Yukawa-like coupling between $\phi$ and $R$, with a small proportionality constant, say $\sigma$.

To this end, we based our analysis on the result found in Ref.~\cite{Dubbini:2025jjz}, where a non-minimal Yukawa-like coupling between the inflaton and gravity, in the context of an expanding universe, was shown to significantly enhance the baryon asymmetry generation with respect to the background. Moreover, we neglect any other possible interaction between the spectator field and the fermions of the Standard Model, assuming them to be suppressed so much so that they can be ignored. In so doing, we can regard $\phi$ as a dark complex scalar field.
We solve the equations of motion for the spectator field, at zero-order in $\xi$ and at first-order in $\sigma$. Then, we compute the inflatonic solution at first-order in $\xi$, neglecting at first stage the expansion of the universe. Afterwards, the baryon asymmetry is thus computed and the effects of the mass-mixing and of the expansion of the universe are included. Our findings, accounting for non-minimal couplings in the inflaton sector, show that possible modification may occur in presence of spectators. Generally, we obtain results that depend on the signs of coupling constants, indicating the regions in which the coupling strengths fail to be predictive since they do not lead to a suitable baryon-antibaryon asymmetry.

The paper is outlined as follows. In Sect.~\ref{sec:level2}, the background model of spontaneous baryogenesis is summarized, emphasizing the broken phase Lagrangian and the result in terms of baryon asymmetry. The model is referred to as zero-order paradigm, that we can obtain by setting $\xi=\sigma=0$. In Sect.~\ref{sec:level3}, the full Lagrangian, containing the background model and the terms involving the spectator field, is presented. The equations of motion of fermions, inflaton, and spectator field, as well as the explicit time-dependent expression for the Ricci scalar are computed. In Sect.~\ref{sec:level4}, first we solve the equation of motion for the spectator field, at zero-order in $\xi$ and at first-order in $\sigma$; thereafter, the equation of motion for the inflaton field is solved at first-order in $\xi$. The solutions are then explained and deeply discussed. In Sect. \ref{sec:level5}, the baryon asymmetry is computed. Finally, in Sect.~\ref{sec:level6}, the effects of the mass-mixing and the expansion of the universe are included. Finally, in Sect.~\ref{sec:level7}, we discuss conclusions and perspectives of our approach.

\section{Spontaneous baryogenesis}\label{sec:level2}

In the framework of spontaneous baryogenesis~\cite{Dolgov:1994zq, Dolgov:1996qq}, the generation of the baryon asymmetry is attributed to baryon-number violating decays of the inflaton field into fermion-antifermion pairs during the reheating phase. In the high-energy regime of the early universe, baryon number is conserved above a characteristic symmetry-breaking scale $f$. Below it, the global $U(1)_B$ symmetry is spontaneously broken, giving rise to a pseudo Nambu-Goldstone boson, which is identified with the inflaton.

During the reheating, the inflaton decays into fermionic species $Q$ and $L$ (and their antiparticles) via baryon-number violating interactions \cite{Belfiglio:2022qai, Belfiglio:2023rxb, Belfiglio:2024swy, Belfiglio:2025chv, Belfiglio:2023moe, Belfiglio:2024xqt}, thereby sourcing a net baryon asymmetry. The theoretical setup assumes a minimal coupling of the inflaton to both gravity and the fermionic degrees of freedom that carry baryon charge. During the broken-symmetry phase, the associated baryonic current,
\begin{equation}
J^\mu = \overline{Q} \gamma^\mu Q,
\end{equation}
is no longer conserved due to the effective interactions emerging from the symmetry breaking, thus enabling the inflaton decays that violate baryon number. This mechanism naturally embeds the conditions for baryogenesis within the post-inflationary dynamics.

\subsection{Initial set up}

In this scenario, the broken phase Lagrangian is
\begin{equation}
\begin{split}
    \mathcal{L}&=\frac{f^2}{2}(\partial_{\mu}\theta)(\partial^{\mu}\theta)-V(\theta)+\overline{Q}(i\gamma^{\mu}\partial_{\mu}-m_Q)Q\\&+\overline{L}(i\gamma^{\mu}\partial_{\mu}-m_L)L+\frac{gf}{\sqrt{2}}(\overline{Q}Le^{i\theta}+\overline{L}Qe^{-i\theta}),
\end{split}
\label{eq1}
\end{equation}
where $f$ is the energy scale at which the symmetry is broken and $g$ is the coupling constant of the interaction between the inflaton field $\theta$ and the fermionic fields, $Q$ and $L$, among which only $Q$ carries a baryon number.

It is worth noting the points summarized below.
\begin{itemize}
    \item[-] The fields $Q$ and $L$ are not real quarks and leptons, because the Lagrangian does not take into account the color charge. Moreover, the inflaton is considered as a classical field, whereas fermions are quantized, following a semiclassical approach.
    \item[-] In Eq.~\eqref{eq1}, $V(\theta)=\Lambda^4[1-\cos(\theta)]$ is the natural inflation potential \cite{Adams:1992bn, Stein:2021uge}. Because the model assumes the baryon asymmetry to be produced during reheating -- when the inflaton oscillates around the minimum of the potential at $\theta=0$ -- we consider a regime of small oscillations that enables a second order expansion for the potential, leading to $V(\theta)\approx m^2f^2\theta^2/2$, with $m=\Lambda^2/f$ defining the bare mass of the inflaton.
    \item[-] Ref.~\cite{Adams:1992bn} constrains (a) the energy scale $f$ to be of the same order of magnitude as the Planck mass $M_{\rm Pl}$, i.e., $f/M_{\text{Pl}}=\mathcal{O}(1)$, and (b) the ratio between $f$ and the amplitude of the potential $\Lambda$ to be $f/\Lambda\sim10^3$--$10^6$. We arbitrary chose to take $f/\Lambda\sim10^3$. This choice does not affect our final result, rather it plays a role in constraining the free parameters to obtain the baryon asymmetry experimentally observed.
\end{itemize}

Eq.~\eqref{eq1} can be rewritten by rotating $Q\rightarrow Q e^{i\theta}$, making the symmetry breaking and the corresponding non-conserved current manifest, i.e.,
\begin{equation}
\begin{split}
    \mathcal{L}&=\frac{f^2}{2}(\partial_{\mu}\theta)(\partial^{\mu}\theta)-V(\theta)+\overline{Q}(i\gamma^{\mu}\partial_{\mu}-m_Q)Q+\\&+\overline{L}(i\gamma^{\mu}\partial_{\mu}-m_L)L+\frac{gf}{\sqrt{2}}(\overline{Q}L+\overline{L}Q)-(\partial_{\mu}\theta) J^{\mu}.
\end{split}
\label{eq2}
\end{equation}
where the baryonic current $J^{\mu}=\overline{Q}\gamma^{\mu}Q$, associated to the global $U(1)$-symmetry, is definitely not conserved in the broken phase, due to the presence of the term $(\partial_{\mu}\theta) J^{\mu}$ that would be canceled out from Eq.~\eqref{eq2}, if the baryon current were conserved, i.e., $\partial_{\mu}J^{\mu}=0$.

The Lagrangian in Eq.~\eqref{eq2} gives rise to the following equation of motion for the inflaton\footnote{The term $3H\dot{\theta}$ does not directly derive from Eq.~\eqref{eq2}. Nevertheless, it is inserted to account for the expansion of the universe.},
\begin{equation}
    \ddot{\theta}+(3H+\Gamma)\dot{\theta}+\Omega^2\theta=0.
\label{eq3}
\end{equation}
where $\Omega$ is the renormalized mass of the inflaton, defined by the formula
\begin{equation}
    m^2=\Omega^2\bigg{[}1+\frac{g^2}{4\pi}\lim_{\omega\to+\infty}\ln\bigg{(}\frac{2\omega}{\Omega}\bigg{)}\bigg{]},
\label{eq4}
\end{equation}
and $\Gamma=g^2\Omega/(8\pi)$ is the decay rate of the inflaton. A simple solution to Eq.~\eqref{eq3} can be found by neglecting the expansion of the universe with respect to the decay rate of the inflaton, i.e., $\Gamma\gg H$. Moreover, the background model assumes the masses of the produced fermions to be much smaller than the mass of the inflaton, constraining the coupling constant to be $g\ll1$ and implying $\Omega\gg\Gamma$.

In this limit, Eq.~\eqref{eq3} has a simple solution
\begin{equation}
    \theta(t)=\theta_Ie^{-\Gamma t/2}\cos(\Omega t),
\label{eq5}
\end{equation}
where $\theta_I$ is the initial value of $\theta$ and refers to the regime in which the expansion of the universe is negligible and it thus coincides with the value of $\theta$ when $H=\Gamma$, that is $\theta_I=\sqrt{3/(4\pi)}\Gamma M_{\text{Pl}}/(f\Omega)$. Supposing that at $t=0$ we have $\theta(t=0)=\theta_I$, then for $t<0$ the inflatonic solution is assumed to be constant, $\theta(t)=\theta_I$.

The solution in Eq.~\eqref{eq5} is then used for computing the baryon asymmetry.
Precisely, the number density $n$ of particle-antiparticle pairs produced by the decay of a homogeneous classical scalar field is computed via
\begin{equation}
    n=\frac{1}{\mathcal V}\sum_{s1,s2}\int\frac{d^3p_1}{(2\pi)^32p_1^0}\frac{d^3p_2}{(2\pi)^32p_2^0}|A|^2,
\label{eq6}
\end{equation}
where $A$ is the single pair production amplitude and $\mathcal V$ the volume element. Using this relation, the number density of baryons is computed by considering the decay of the inflaton in $Q-\overline{L}$ pairs, whereas the number density of antibaryons by the decay of the inflaton in $\overline{Q}-L$ pairs. The corresponding decay amplitudes are thus computed. The baryon asymmetry $n_B$ is finally figured out by taking the difference between the number density of baryons and antibaryons, namely
\begin{equation}
    n_B=\frac{1}{16\pi}\Omega g^2f^2\theta_I^3.
\label{eq7}
\end{equation}
Eq.~\eqref{eq7} takes into account neither the expansion of the universe nor the mass-mixing. Once included, they enhance Eq.~\eqref{eq7} by factors $8\pi/g^2$ and $[(1-\epsilon^2)/(1+\epsilon^2)]^2$, respectively, where $\epsilon=\sqrt{2}gf/[\Delta m+\sqrt{\Delta m^2+2g^2f^2}]$ depends upon the difference $\Delta m=m_Q-m_L$ of the masses of fermions $Q$ and $L$ \cite{Dolgov:1996qq}.

\section{Effects of a  spectator field within spontaneous baryogenesis}\label{sec:level3}

Here, we extend the background model of spontaneous baryogenesis by introducing a subdominant complex scalar spectator field, $\phi$, weakly coupled to the inflaton.
To do so, we consider two main interactions, as below reported.

\begin{itemize}
    \item[-] A bilinear quadratic coupling $\propto \theta^2|\phi|^2$, aiming at obtaining a correction to the inflaton mass and, consequently, to modify its decay amplitude into fermion-antifermion pairs, possibly leading to an enhancement of the baryon asymmetry production.
    \item[-] A non-minimal Yukawa-like coupling between the spectator field and gravity via $\propto R|\phi|^2$, as motivated by recent studied showing how the baryon asymmetry can be increased accordingly.
\end{itemize}

Any other plausible interaction between $\phi$ and the fermions of the Standard Model is here ignored, considering \emph{de facto} the spectator just as a dark scalar field.

To work the two above-mentioned options out, we employ a  spatially flat Friedmann-Robertson-Walker (FRW) universe, generalizing Eq.~\eqref{eq2} to
\begin{equation}
\begin{split}
    \mathcal{L}_0&=\frac{f^2}{2}(\nabla_{\mu}\theta)(\nabla^{\mu}\theta)-V(\theta)+\\&+\frac{i}{2}\bigg{(}\overline{Q}\gamma^{\alpha}\nabla_{\alpha}Q-\nabla_{\alpha}(\overline{Q}\gamma^{\alpha})Q\bigg{)}-m_Q\overline{Q}Q+\\&+\frac{i}{2}\bigg{(}\overline{L}\gamma^{\alpha}\nabla_{\alpha}L-\nabla_{\alpha}(\overline{L}\gamma^{\alpha})L\bigg{)}-m_L\overline{L}L+\\&+\frac{gf}{\sqrt{2}}(\overline{Q}L+\overline{L}Q)-(\nabla_{\mu}\theta) J^{\mu},
\end{split}
\label{eq8}
\end{equation}
where the covariant derivatives, in lieu of the partial derivatives for both for the inflaton and fermions, appear as a consequence of the fact that our quantities are computed in curved space, i.e., since we consider the gravitational field non-minimally coupled to the spectator.
For this reason, knowing the time-dependent expression for $R$ is strictly necessary to determine the equations of motion. In this context, therefore, just adding the Hubble-friction term in the equations of motion as in the background is not enough to account for the curvature of the spacetime. Conversely, the full covariant formalism of curved spaces should be used and, in particular, the covariant derivatives for the fields must be properly defined.

For the inflaton field, $\theta$, the covariant derivative coincides with the partial derivative, since the inflaton is a scalar field. In contrast, in the case of fermions, the covariant derivatives are defined through the tetrad formalism. In particular, working in a spatially flat FRW spacetime, with metric
\begin{equation}
g_{\mu\nu}=dt^2-a^2(t)(dx^2+dy^2+dz^2),
\label{eq9}
\end{equation}
we adopt the following choice of tetrads,
\begin{equation}
    e^a_{\mu}=
    \begin{pmatrix}
    1 & 0 & 0 & 0 \\
    0 & a & 0 & 0 \\
    0 & 0 & a & 0 \\
    0 & 0 & 0 & a
    \end{pmatrix}.
\label{eq10}
\end{equation}
that satisfies the conditions $g_{\mu\nu}=\eta_{ab}e^a_{\mu}e^b_{\nu}$ and $\eta_{ab}=g_{\mu\nu}e^{\mu}_ae^{\nu}_b$, where $\eta_{\mu\nu}$ is the Minkowski metric, and $e^{\mu}_a$ is the inverse matrix of $e^a_{\mu}$.
Having chosen the tetrads, we can write the gamma matrices in curved spacetimes as $\gamma^{\mu}=e^{\mu}_a\gamma^a$, where $\gamma^a$ represent the Dirac matrices in Minkowski spacetime. Furthermore, the covariant derivative of a spinor $\psi$ can be written as
\begin{equation}
\nabla_{\mu}\psi=\partial_{\mu}\psi+\chi_{\mu}\psi=\partial_{\mu}\psi+\frac{1}{8}\omega_{\mu b c}[\gamma^b,\gamma^c]\psi,
\label{eq11}
\end{equation}
where the spin connection $\omega_{\mu b c}$ is given by
\begin{equation}
\omega_{\mu a b}=\eta_{ac}e^c_{\nu}e^{\sigma}_b\Gamma^{\nu}_{\sigma\mu}+\eta_{ac}e^c_{\nu}\partial_{\mu}e^{\nu}_b,
\label{eq12}
\end{equation}
and, with our tetrads, we have $\chi_0=0$ and $\chi_i=\dot{a}\gamma_i\gamma_0/2$.

The generalization to the curved FRW spacetime can be finally recovered by adding to the Lagrangian $\mathcal{L}_0$ the Einstein-Hilbert term for the gravitational field, that is $-M^2_{\text{Pl}}R/(16\pi)$, yielding to
\begin{equation}
\mathcal{L}_{\text{bkg}}=-\frac{M^2_{\text{Pl}}}{16\pi}R+\mathcal{L}_0.
\label{eq12a}
\end{equation}

As mentioned above, we now include the Lagrangian of a complex scalar spectator field $\phi$, weakly coupled to $\theta$ and non-minimally coupled to gravity, through a Yukawa-like coupling with the Ricci curvature
\begin{equation}
\mathcal{L}_{\phi}=(\nabla^{\alpha}\phi^*)(\nabla_{\alpha}\phi)-m^2_{\phi}\phi^*\phi-\xi f^2\theta^2\phi^*\phi-\sigma R\phi^*\phi.
\label{eq13}
\end{equation}
The complex scalar field is defined by means of two real, non-interacting scalar fields, namely
\begin{equation}
\phi=\frac{\phi_1+i\phi_2}{\sqrt{2}}.
\label{eq13a}
\end{equation}
With this position, the Lagrangian in Eq.~\eqref{eq13} becomes the sum of two independent real scalar field Lagrangians
\begin{equation}
\mathcal{L}_{\phi}=\mathcal{L}_{\phi_1}+\mathcal{L}_{\phi_2}
\label{eq13c}
\end{equation}
where $\mathcal{L}_{\phi_1}$ and $\mathcal{L}_{\phi_2}$, respectively, referring to the real fields $\phi_1$ and $\phi_2$, are defined as
\begin{equation}
\mathcal{L}_{\phi_{1,2}}\!=\!\frac{1}{2}\nabla^{\alpha}\phi_{1,2} \nabla_{\alpha}\phi_{1,2}-\frac{1}{2}(m^2_{\phi}\!+\!\xi f^2\theta^2\!+\!\sigma R)\phi^2_{1,2}.
\label{eq13b}
\end{equation}
Hence, the total Lagrangian under exam is given by
\begin{equation}
\mathcal{L}=\mathcal{L}_{bkg}+\mathcal{L}_{\phi_1}+\mathcal{L}_{\phi_2},
\label{eq14}
\end{equation}
leading to the equation of motion for the inflaton
\begin{equation}
f^2\nabla^2\theta+V'(\theta)+\xi f^2(\phi_1^2+\phi_2^2)\theta=\braket{\nabla_{\mu}J^{\mu}},
\label{eq15}
\end{equation}
that in a FRW universe becomes
\begin{equation}
f^2(\ddot{\theta}+3H\dot{\theta})+f^2[m^2+\xi(\phi_1^2+\phi_2^2)]\theta=\braket{\nabla_{\mu}J^{\mu}}.
\label{eq16}
\end{equation}

To solve Eq.~\eqref{eq16} we require the time-dependent expression for the spectator field and the vacuum expectation value (VEV) of the baryonic current $\braket{\nabla_{\mu}J^{\mu}}$. The latter can be conveniently written as
\begin{equation}
\braket{\nabla_{\mu}J^{\mu}}=\frac{igf}{\sqrt{2}}\braket{\overline{Q}L-\overline{L}Q}.
\label{eq17}
\end{equation}

Thus, to compute it, we have to solve the equations of motion for the fermionic fields $Q$ and $L$, respectively
\begin{subequations}
    \begin{align}
    &i\gamma^{\mu}\nabla_{\mu}Q-m_QQ-(\nabla_{\mu}\theta)\gamma^{\mu}Q=-\frac{gf}{\sqrt{2}}L,\label{eq18}\\
    &i\gamma^{\mu}\nabla_{\mu}L-m_LL=-\frac{gf}{\sqrt{2}}Q,\label{eq19}
    \end{align}
\end{subequations}
becoming in a FRW universe\footnote{In an FRW universe, classical fields depend only on cosmic time to preserve homogeneity and isotropy.
Quantized fields like fermions are functions of a point in the spacetime. So, the gradient term in the equations of motion cannot be neglected.}
\begin{subequations}
    \begin{align}
    &\gamma^0\!\left(\!\dot{Q}\!+\!\frac{3H}{2}Q\!\right)\!+\!\frac{\gamma^i}{a}\partial_iQ\!+i (m_Q \!+\! \dot{\theta}\gamma^0)Q\!=\!\frac{igf}{\sqrt{2}}L,\label{eq20}\\
    &\gamma^0\bigg{(}\dot{L}+\frac{3H}{2}L\bigg{)}+\frac{\gamma^i}{a}\partial_iL+i m_LL = \frac{igf}{\sqrt{2}}Q.
\label{eq21}
    \end{align}
\end{subequations}

The equations of motion for $Q$ and $L$ can be solved at first order in perturbation theory assuming $g\ll1$.

\subsection{Perturbative computation}

At this stage, we neglect the expansion of the universe, that will be restored later, assuming a regime in which the scale factor $a(t)$ slowly varies over the cosmic time and, thus, it can be approximated with $a(t)\simeq\text{const}$.

At zero order, the equations of motion for $L$ and $\tilde{Q}$, defined by $Q=\tilde{Q}e^{-i\theta}$, reduce to the free Dirac equations in the FRW spacetime.
The overall computation of the FRW free Dirac equation, with $a\simeq \text{const}$, is reported in Appendix \ref{app1}.
Thus, the corresponding zero-order solutions $Q_0$ and $L_0$ are
\begin{subequations}
\begin{align}
\nonumber
&Q_0(t,\vec{x})=a^{-\frac{3}{2}}(t)e^{-i\theta(t)}\sum_s\int\frac{d^3p}{(2\pi)^3}\frac{1}{\sqrt{2E_p}}\times \\
\label{eq22}
&\!\left[u(\vec{p},\!s)\hat a(\vec{p},\!s)e^{-i(\!E_pt-\vec{p}\cdot\vec{x})}\!+\!v(\vec{p},\!s)\hat b^{\dag}\!(\vec{p},\!s)e^{i(\!E_pt-\vec{p}\cdot\vec{x})}\!\right]\!\\
\nonumber
&L_0(t,\vec{x})=a^{-\frac{3}{2}}(t)\sum_s\int\frac{d^3p}{(2\pi)^3}\frac{1}{\sqrt{2E_p}}\times \\
&\!\left[u(\vec{p},\!s)\hat c(\vec{p},\!s)e^{-i(\!E_pt-\vec{p}\cdot\vec{x})}\!+\!v(\vec{p},\!s)\hat d^{\dag}\!(\vec{p},\!s)e^{i(\!E_pt-\vec{p}\cdot\vec{x})}\!\right]\!
\label{eq23}
\end{align}
\end{subequations}
where $\vec{p}$ is the three-momentum, $E_p$ is the energy, and $u$ and $v$ are spinors. Further, we indicate with $\hat a$ and $\hat c$ the annihilation operators, whereas with $\hat b^\dagger$ and $\hat d^\dagger$ the creation operators.
The solutions Eqs.~\eqref{eq22}-\eqref{eq23}
differ from those in the Minkowski spacetime just by a factor $a^{-3/2}$, which is reasonable since the modification relates with $\mathcal V^{1/2}$ that, in the FRW universe, scales as $\mathcal V\propto a^3$.

The first-order solutions $Q_1$ and $L_1$ of Eqs.~\eqref{eq21}--\eqref{eq22} are, respectively,
\begin{subequations}
\begin{align}
\nonumber
&Q_1(t,\vec{x})=Q_0(t,\vec{x})-a^{-\frac{3}{2}}(t)e^{-i\theta(t)}\frac{gf}{\sqrt{2}}\times \\
&\times \int dt'\int d^3y G_Q(t,\vec{x};t',\vec{y})L_0(t',\vec{y})a^{\frac{3}{2}}(t')e^{i\theta(t')}, \label{eq24}\\
\nonumber
&L_1(t,\vec{x})=L_0(t,\vec{x})-a^{-\frac{3}{2}}(t)\frac{gf}{\sqrt{2}}\times \\& \times \int dt'\int d^3y G_L(t,\vec{x};t',\vec{y})Q_0(t',\vec{y})a^{\frac{3}{2}}(t').
\label{eq25}
\end{align}
\end{subequations}
Following Ref.~\cite{Dolgov:1994zq}, the VEV of $\nabla_{\mu}J^{\mu}$ in Eq.~\eqref{eq17} coincides with the results of the background model corrected by a factor $a^{-3}$, as expected for the FRW spacetime when considering $a$ slowly varying with the cosmic time,
\begin{equation}
\braket{\nabla_{\mu}J^{\mu}}=-
\frac{g^2f^2\Omega}{4\pi\,a^3} \left[\frac{\dot{\theta}}{2}-\Omega\theta \lim_{\omega\to+\infty}\ln\bigg{(}\frac{2\omega}{\Omega}\bigg{)}\right].
\label{eq26}
\end{equation}
Considering that at this stage we are assuming $a\simeq\text{const}$, for simplicity we can set $a\simeq 1$, thus, recovering the same result as in the background model.
Moreover, since fermions are not affected by the presence of the spectator field,  $\braket{\nabla_{\mu}J^{\mu}}$ is independent from $\xi$ and $\sigma$.

Using the VEV of Eq.~\eqref{eq26} and recalling that neglecting the expansion of the universe implies $H\ll\Gamma$, the equation of motion for the inflaton in Eq.~\eqref{eq16} becomes
\begin{equation}
\ddot{\theta}+\Gamma\dot{\theta}+(\Omega^2+\xi \phi_1^2+\xi\phi_2^2)\theta=0,
\label{eq28}
\end{equation}
Looking at Eq.~\eqref{eq28}, it is evident that the inclusion of a complex scalar spectator field $\phi$, coupled with the inflaton, apports a first-order correction $\xi(\phi_1^2+\phi_2^2)$ to the mass of the inflaton itself. Since $\phi_1^2+\phi_2^2$ is positive, the corrective term sign completely depends on the sign of $\xi$.

In particular, to enhance baryogenesis, we definitely need an increase of inflaton mass, thus increasing the probability of inflaton decay into fermion-antifermion pairs. This implies that, in order to have a positive correction to the mass of the inflaton, we end up with $\xi>0$.

Since the inflaton and the spectator fields are coupled, solving Eq.~\eqref{eq28} requires the corresponding equations of motion for the real part, $\phi_1$, and imaginary part, $\phi_2$, of the complex spectator field, respectively,
\begin{equation}
\nabla^2\phi_{1,2}+(m^2_{\phi}+\xi f^2\theta^2+\sigma R)\phi_{1,2}=0,
\label{eq29a}
\end{equation}
becoming in the FRW spacetime,
\begin{equation}
\ddot{\phi}_{1,2}+3H\dot{\phi}_{1,2}+(m^2_{\phi}+\xi f^2\theta^2+\sigma R)\phi_{1,2}=0.
\label{eq30a}
\end{equation}

\subsection{Solving the equations of motion}

We immediately notice that $\phi_1$ and $\phi_2$ satisfy the same equations of motion, having no interactions with each other. Thus, from now on, we consider $\phi=\phi_1=\phi_2$, where $\phi$ shall not be confused with the spectator field, but shall be only identified with its real and imaginary parts.
In so doing, for the inflaton and the spectator fields we obtain, respectively
\begin{subequations}
\begin{align}
&\ddot{\theta}+\Gamma\dot{\theta}+(\Omega^2+2\xi \phi^2)\theta=0,
\label{eq30b}\\
&\ddot{\phi}+3H\dot{\phi}+(m^2_{\phi}+\xi f^2\theta^2+\sigma R)\phi=0.
\label{eq30c}
\end{align}
\end{subequations}
We can now proceed analogously to inflationary equation of motion, by neglecting the Hubble-friction term. However, for the equation of motion of the inflaton, it was sufficient to assume $\Gamma\gg H$ in order to neglect the universe expansion. Conversely, here, there is no other term proportional to the first derivative of the field to be compared to the Hubble-friction term.

To better understand this point, we consider Eq.~\eqref{eq30c} at zero-order, say for $\xi=\sigma=0$, and neglect the term $\propto H\dot\phi$. It is immediate to see that it describes a harmonic oscillator with frequency $m_{\phi}$, with all terms of  $m^2_{\phi}\phi$ order. Consequently, the Hubble friction term would be of the order of $m_{\phi}H\phi$, so neglecting it bounds $H\ll m_{\phi}$.

Given that we assume $\Gamma \gg H$, the condition $H \ll m_{\phi}$ is automatically fulfilled provided that $m_{\phi} > \Gamma$, which is a physically reasonable assumption. Indeed, the mass of the spectator field is expected to be smaller than inflaton masses, yet not entirely negligible. In particular, the inequality $\Omega \gg \Gamma$ typically holds, ensuring that the spectator dynamics remain distinct from the Hubble scale and the decay rate of the inflaton.

Solving Eq.~\eqref{eq30c} also requires finding a suitable time-dependent expression for the Ricci scalar curvature $R$. To this end, we write down the equation of motion for the gravitational field, accounting for $\phi=\phi_1=\phi_2$,
\begin{equation}
\begin{split}
&\frac{M^2_{\text{Pl}}}{8\pi}\left(R_{\mu\nu}-\frac{1}{2}g_{\mu\nu}R \right)+\\
&\sigma \left(R_{\mu\nu}-\frac{1}{2}g_{\mu\nu}R + g_{\mu\nu}\nabla^2-\nabla_{\mu}\nabla_{\nu}\right) 2\phi^2=T_{\mu\nu},
\end{split}
\label{eqGRAV1}
\end{equation}
in which $T_{\mu\nu}$ is the stress-energy tensor associated with the matter Lagrangian, including the spectator field, whose full expression and trace are reported in Appendix~\ref{app1}, respectively in Eqs.~\eqref{eqENIMPTENSOR}--\eqref{eqENIMPTENSORTRACE}.
By working the trace of Eq.~\eqref{eqGRAV1} out, we thus obtain
\begin{equation}
-\frac{M^2_{\text{Pl}}}{8\pi}R-\sigma(R-3\nabla^2)2\phi^2=T^{\mu}_{\mu},
\label{eqGRAV2}
\end{equation}
where $T^{\mu}_{\mu}$ is the trace of the stress-energy tensor that, using Eqs.~\eqref{eq18}--\eqref{eq19}, further simplifies into
\begin{align}
\nonumber
&T^{\mu}_{\mu}=-f^2(\nabla^{\alpha}\theta)(\nabla_{\alpha}\theta)+4V(\theta)+m_Q\overline{Q}Q+m_L\overline{L}L+\\
\nonumber
&-(\nabla_{\alpha}\theta)J^{\alpha}-\frac{gf}{\sqrt{2}}(\overline{Q}L+\overline{L}Q)+\\
\label{eqENIMPTENSORTRACEa}
&-2(\nabla^{\alpha}\phi)(\nabla_{\alpha}\phi) + 4m^2_{\phi}\phi^2+4\xi f^2\theta^2\phi^2.
\end{align}
Finally, Eq.~\eqref{eqGRAV2} yields
\begin{align}
\nonumber
&-\frac{M^2_{\text{Pl}}}{8\pi}R-\sigma(R-3\nabla^2)2\phi^2=-f^2(\nabla^{\alpha}\theta)(\nabla_{\alpha}\theta)+\\
\nonumber
&m_Q\overline{Q}Q+m_L\overline{L}L-(\nabla_{\alpha}\theta)J^{\alpha}-\frac{gf}{\sqrt{2}}(\overline{Q}L+\overline{L}Q)+\\
\label{eqGRAV3}
&4V(\theta)-2(\nabla^{\alpha}\phi)(\nabla_{\alpha}\phi)+4m^2_{\phi}\phi^2+4\xi f^2\theta^2\phi^2.
\end{align}

Since $g\ll1$, we expect the fermion-antifermion couples produced by the decays of the inflaton to prompt very small masses, roughly of order $\sim gf$. Thus, we may handle the approximation of massless fermions, holding as long as $m_Q,m_L\ll\Omega$, without altering the solution for $\theta$ significantly \cite{Dolgov:1996qq}.
In this regime, in a spatially flat FRW spacetime, Eq.~\eqref{eqGRAV3} reads
\begin{equation}
\begin{split}
&\bigg{(}\frac{M^2_{\text{Pl}}}{8\pi}+2\sigma\phi^2\bigg{)}R-12\sigma(\dot{\phi}^2+\phi\ddot{\phi}+3H\phi\dot{\phi})=\\&=f^2\dot{\theta}^2-4V(\theta)+\dot{\theta}\braket{J^0}+\frac{gf}{\sqrt{2}}\braket{\overline{Q}L+\overline{L}Q}\\&+2\dot{\phi}^2-4m^2_{\phi}\phi^2-4\xi f^2\theta^2\phi^2.
\end{split}
\label{eqGRAV4}
\end{equation}

In computing the VEVs in Eq.~\eqref{eqGRAV4}, we once again use perturbation theory. In particular, for $\braket{\overline{Q}L+\overline{L}Q}$, we use perturbation theory at first order, finding
\begin{equation}
\frac{gf}{\sqrt{2}}\braket{\overline{Q}L+\overline{L}Q}=a^{-3}(t)\frac{g^2f^2\Omega^2}{8\pi}\theta(t),
\label{eqVEV1}
\end{equation}
whereas for $\braket{J^0}$ working at zero-order in perturbation theory is enough to obtain
\begin{equation}
\braket{J^0}=\frac{P^3_{\text{max}}}{3\pi^2}a^{-3}(t),
\label{eqVEV2}
\end{equation}
in which $P_{\text{max}}$ is a cutoff scale, introduced to avoid the natural divergence arising from computing energy density in quantum field theories. Hereafter, $P_{\text{max}}$ turns out to be a free parameter of the theory. Roughly, we do expect it to lie on the energy scale of $f$. It is important to remark that the expressions for the above-mentioned VEVs are valid just if the solutions for the Dirac fields are those reported in Appendix~\ref{app1}, assuming thus $a(t)\approx \text{const}$.

Eq.~\eqref{eqGRAV4} can be greatly simplified at zero order:
\begin{itemize}
    \item[-] for $\xi=0$, to avoid an expression for the spectator field at the first order in $\xi$ that would introduce at least a term $\propto\xi^2$ in the equation of motion for the inflaton, that is negligible;
    \item[-] for $\sigma=0$, because in the spectator equation of motion $R$ is already multiplied by $\sigma$, hence, considering terms at first-order in $\sigma$ in Eq.~\eqref{eqGRAV4} would introduce unnecessary second-order terms in $\sigma$.
\end{itemize}
Thus, in view of the above considerations and working in the regime at which $a(t)\approx1$, the equation of motion for the gravitational field in Eq.~\eqref{eqGRAV4} can be simplified to
\begin{equation}
\begin{split}
&\frac{M^2_{\text{Pl}}}{8\pi}R=f^2\dot{\theta}^2-4V(\theta)+2\dot{\phi}^2-4m^2_{\phi}\phi^2+\\&+\frac{P^3_{\text{max}}}{3\pi^2}\dot{\theta}+\frac{g^2f^2\Omega^2}{8\pi}\theta.
\end{split}
\label{eqGRAV6}
\end{equation}
Recalling that $P_{\text{max}}\sim f$, $\Omega\sim 10^{-6}f$ and, coherently with the background model, $\theta\sim\theta_I\sim g^2$ and $\dot{\theta}\sim\Omega\theta$, we easily prove that the dominant term of Eq.~\eqref{eqGRAV6}, among those containing the inflatonic solution, is the one proportional to $P^3_{\text{max}}$. Therefore, Eq.~\eqref{eqGRAV6} simplifies to
\begin{equation}
\begin{split}
&\frac{M^2_{\text{Pl}}}{8\pi}R=2\dot{\phi}^2-4m^2_{\phi}\phi^2+\frac{P^3_{\text{max}}}{3\pi^2}\dot{\theta}.
\end{split}
\label{eqGRAVNEW}
\end{equation}
In addition, the field $\phi$ is a spectator and, then, subdominant with respect to the inflaton, at least during the its decay into fermion-antifermion pairs, therefore its contribution to the stress-energy tensor is negligible.
This holds for $f^3\Omega \theta_I\gg m^2_{\phi}\phi^2_I$, with $\phi_I$  the initial value of the spectator field. In this respect, the most recent Planck's measurements show that the upper bound for the inflatonic scale is represented by the GUT scale, meaning that we can consider $\phi_I\sim 10^{16}\text{ GeV}\sim 10^{-3}f$. The above-mentioned constraint can be thus rewritten as
\begin{equation}
\theta_I\gg \frac{m^2_{\phi}}{f^2}\implies g\gg\frac{m_{\phi}}{f}.
\label{eqANOTHEREQ}
\end{equation}

\begin{figure*}
\centering
\includegraphics[width=0.9\hsize,clip]{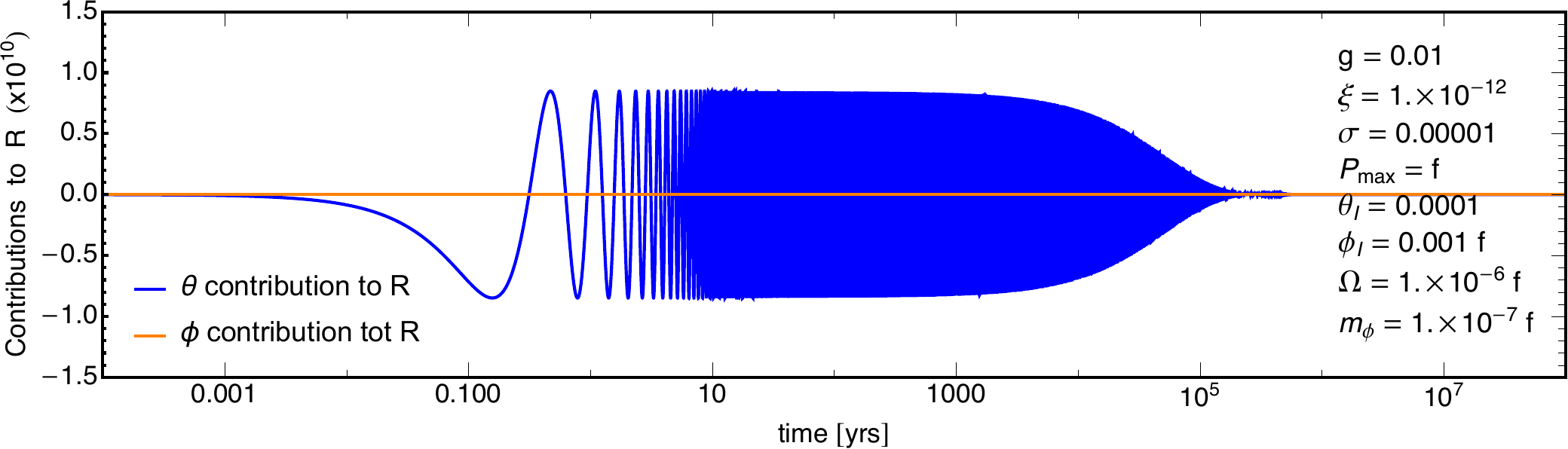}\\
\caption{Comparison between the contributions to the Ricci scalar of inflaton and spectator fields from Eq.~\eqref{eqGRAVNEW}. The parameters utilized for the plot are $g=0.01$, $\Omega=10^{-6}f$, $\xi=10^{-12}$, $\sigma=10^{-5}$, $\phi_I=10^{-3}f$ and $m_{\phi}=10^{-7}f$, for which the condition $g\gg m_{\phi}/f$ holds and, thus, the contribution of the spectator field is negligible.}
\label{fig1}
\end{figure*}

Fig.~\ref{fig1} portrays the separate contributions of spectator and inflaton fields to the full expression of the Ricci scalar from Eq.~\eqref{eqGRAVNEW}, confirming that neglecting the contribution of the spectator is licit as long as $g\gg m_{\phi}/f$ holds.
Thus, Eq.~\eqref{eqGRAVNEW} reduces to
\begin{equation}
R=\Upsilon\dot{\theta},\qquad \Upsilon=\frac{8\pi}{M^2_{\text{Pl}}}\frac{P^3_{\text{max}}}{3\pi^2},
\label{eqRICCI}
\end{equation}
where the parameter $\Upsilon$ has the dimension of an energy and is of the order of $f$.

Now, we can finally rewrite Eq.~\eqref{eq30c}, using the time-dependent expression of $R$ from Eq.~\eqref{eqRICCI}, and neglecting the expansion of the universe via the condition $m_{\phi}>\Gamma$. The so-obtained equation of motion for the spectator, together with Eq.~\eqref{eq30b} for the inflaton, form a second order differential equation system,
\begin{equation}
\begin{cases}
\ddot{\theta}+\Gamma\dot{\theta}+(\Omega^2+2\xi \phi^2)\theta=0,\\
\ddot{\phi}+(m^2_{\phi}+\xi f^2\theta^2+\sigma\Upsilon\dot{\theta})\phi=0,
\end{cases}
\label{eqEOMs}
\end{equation}
yielding the following initial conditions,
\begin{equation}
\begin{cases}
\theta(0)=\theta_I, \ \ \ \dot{\theta}(0)=0,\\
\phi(0)=\phi_I, \ \ \ \dot{\phi}(0)=0,
\end{cases}
\label{eqINCOND}
\end{equation}
which must be jointly solved in order to obtain the dynamics of both spectator and inflaton fields.

\section{Solving the inflatonic dynamics}\label{sec:level4}

We here perturbatively solve in $\xi$ the system~\eqref{eqEOMs} supposing the following inflaton dynamics,
\begin{equation}
\theta(t)=\theta_{0\xi}(t)+\xi\theta_{1\xi}(t),
\label{eqEOM1}
\end{equation}
where $\theta_{0\xi}(t)$ is the  background solution in Eq.~\eqref{eq5}, and $\theta_{1\xi}(t)$ is the first-order contribution in $\xi$.

To determine the time-dependent solution for the spectator field in the perturbative regime at first-order, we have to deal with both the coupling constants $\xi$ and $\sigma$. However, considering a solution for the spectator field up to first-order in $\xi$ would introduce -- through the term $2\xi\phi^2\theta$ -- at least second-order terms in $\xi$ in the inflatonic equation of motion. Therefore, the solution for the spectator must be at zero-order in $\xi$ and, for simplicity, we assume the following form
\begin{equation}
\phi(t)=\phi_{0\xi}(t)+\xi\phi_{1\xi}(t).
\label{eqEOM2}
\end{equation}
where again $\phi_{0\xi}(t)$ and $\phi_{1\xi}(t)$ label, respectively, the zero-order and the first-order contributions in $\xi$.

Therefore, using Eqs.~\eqref{eqEOM1}--\eqref{eqEOM2} respectively in lieu of $\theta$ and $\phi$ and keeping only terms up to the first-order in $\xi$, the equation of motion for the inflaton becomes
\begin{equation}
\begin{split}
&\ddot{\theta}_{0\xi}+\Gamma\dot{\theta}_{0\xi}+\Omega^2\theta_{0\xi}+\\&+\xi(\ddot{\theta}_{1\xi}+\Gamma\dot{\theta}_{1\xi}+\Omega^2\theta_{1\xi}+2\phi_{0\xi}^2\theta_{0\xi})=0.
\end{split}
\label{eqEOMINFL1}
\end{equation}
where the zero-order equation of motion reduces to the background model, guaranteeing that $\theta_{0\xi}(t)$ effectively coincides with Eq.~\eqref{eq5}.

Now, using again Eqs.~\eqref{eqEOM1}--\eqref{eqEOM2} and retaining only zero-order terms in $\xi$ and the zero-order solution $\phi_{0\xi}(t)$, the resulting equation of motion for the spectator is
\begin{equation}
\begin{split}
&\ddot{\phi}_{0\xi}+m^2_{\phi}\phi_{0\xi}+\sigma\Upsilon\dot{\theta}_{0\xi}\phi_{0\xi}=0,
\end{split}
\label{eqEOM4}
\end{equation}
where $\dot{\theta}_{0\xi}(t)$, in the limit $\Omega\gg\Gamma$, is given by
\begin{equation}
\dot{\theta}_{0\xi}(t)\simeq -\Omega\theta_Ie^{-\Gamma t/2}\sin(\Omega t).
\label{eqEOM5}
\end{equation}

Now, we proceed to solve Eq.~\eqref{eqEOM4} for the spectator field working in perturbative regime at first-order of the coupling constant $\sigma$ associated to the non-minimal Yukawa-like coupling  between $\phi$ and gravity. For this reason, we split $\phi_{0\xi}(t)$ in two contributions
\begin{equation}
\phi_{0\xi}(t)=\phi_{0\xi,0\sigma}(t)+\sigma\phi_{0\xi,1\sigma}(t),
\label{eqEOM6}
\end{equation}
where $\phi_{0\xi,0\sigma}(t)$ and $\phi_{0\xi,1\sigma}(t)$ label, respectively, the zero-order and the first-order contributions in $\sigma$.
Inserting Eq.~\eqref{eqEOM6}, in lieu of $\phi_{0\xi}(t)$, into Eq.~\eqref{eqEOM4} provides
\begin{equation}
\begin{split}
&\ddot{\phi}_{0\xi,0\sigma}+\sigma\ddot{\phi}_{0\xi,1\sigma}+m^2_{\phi}(\phi_{0\xi,0\sigma}+\sigma\phi_{0\xi,1\sigma})+\\&+\sigma\Upsilon\dot{\theta}_{0\xi}(\phi_{0\xi,0\sigma}+\sigma\phi_{0\xi,1\sigma})=0,
\end{split}
\label{eqEOM7}
\end{equation}
that can be perturbatively split into the equations of motion for $\phi_{0\xi,0\sigma}(t)$ and $\phi_{0\xi,1\sigma}(t)$, respectively
\begin{subequations}
\begin{align}
&\ddot{\phi}_{0\xi,0\sigma}+m^2_{\phi}\phi_{0\xi,0\sigma}=0,
\label{eqEOM8}\\
&\ddot{\phi}_{0\xi,1\sigma}+m^2_{\phi}\phi_{0\xi,1\sigma}+\Upsilon\dot{\theta}_{0\xi}\phi_{0\xi,0\sigma}=0.
\label{eqEOM9}
\end{align}
\end{subequations}
The solution of Eq.~\eqref{eqEOM8} describes a harmonic oscillator with initial conditions $\phi_{0\xi,0\sigma}(0)=\phi_I$ and $\dot{\phi}_{0\xi,0\sigma}(0)=0$, extending \emph{de facto} the ones in Eq.~\eqref{eqINCOND}
\begin{equation}
\phi_{0\xi,0\sigma}(t)=\phi_I\cos(m_{\phi}t).
\label{eqEOM10}
\end{equation}
Using Eq.~\eqref{eqEOM10}, Eq.~\eqref{eqEOM9} for $\phi_{0\xi,1\sigma}(t)$ becomes
\begin{equation}
\ddot{\phi}_{0\xi,1\sigma}\!+\!m^2_{\phi}\phi_{0\xi,1\sigma}\!=\!\Upsilon\Omega\theta_I\phi_Ie^{-\Gamma t/2}\sin(\Omega t)\cos(m_{\phi}t),
\label{eqEOM11}
\end{equation}
for which, imposing as initial conditions $\phi_{0\xi,1\sigma}(0)=0$ and $\dot{\phi}_{0\xi,1\sigma}(0)=0$, we find the following solution
\begin{equation}
\phi_{0\xi,1\sigma}(t)\!=\!\Upsilon\theta_I\phi_I\!\left[\frac{ \cos(m_\phi t)\sin(\Omega t)}{\Omega e^{\Gamma t/2}}\!-\!\frac{\sin(m_\phi t)}{m_\phi}\right].
\label{eqEOM12}
\end{equation}
Putting together Eqs.~\eqref{eqEOM10} and \eqref{eqEOM12}, we obtain the full time-dependent solution $\phi_{0\xi}(t)$
\begin{align}
\nonumber
\phi_{0\xi}(t)=\phi_I\Big\{&\cos(\alpha\Omega t)-\Sigma\Big[\sin(\alpha\Omega t)+\\
&\left.\left.-\alpha e^{-\Gamma t/2}\sin(\Omega t)\cos(\alpha\Omega t)\right]\right\}.
\label{eqEOM15}
\end{align}
where, for simplicity, we defined the dimensionless parameters $\Sigma=\sigma\Upsilon\theta_I/m_{\phi}$ and $\alpha=m_{\phi}/\Omega$.

Since we are working out a perturbation method, the first-order correction of Eq.~\eqref{eqEOM15} has to be smaller than the zero-order solution, which is equivalent to require
\begin{equation}
\Sigma\ll1\implies \sigma\ll 10^{-6}\frac{m_{\phi}}{\Gamma}\sim \frac{1}{g^2}\frac{m_{\phi}}{f}.
\label{eqEOM16}
\end{equation}
Moreover, since the mass of the spectator is expected to be smaller than the mass of the inflaton, or equivalently $\alpha<1$, definitely also the condition $\alpha\Sigma\ll1$ shall hold.

Now, back to the solution of the inflatonic first-order equation of motion, Eq.~\eqref{eqEOMINFL1} and can be rewritten using Eqs.~\eqref{eqEOM15} and \eqref{eq5} in lieu respectively of $\phi_{0\xi}(t)$ and $\theta_{0\xi}(t)$,
and keeping only terms up to first-order in $\Sigma$

\begin{align}
\nonumber
&\ddot{\theta}_{1\xi}+\Gamma\ddot{\theta}_{1\xi}+\Omega^2\theta_{1\xi}=-2\theta_I\phi_I^2e^{-\Gamma t/2}\cos(\Omega t)\cos^2(\alpha\Omega t)\times\\
&\times \bigg{[}1-2\Sigma\tan(\alpha\Omega t)+2\alpha\Sigma e^{-\Gamma t/2}\sin(\Omega t)\bigg{]}.
\label{eqEOM18}
\end{align}
To solve Eq.~\eqref{eqEOM18} we impose the initial conditions $\theta_{1\xi}(0)=0$ and $\dot{\theta}_{1\xi}(0)=0$, extending \emph{de facto} those in Eq.~\eqref{eqINCOND}; further, we work out the limit $\Omega\gg\Gamma$ and neglect terms proportional to $\alpha^2$, but not those proportional to $\alpha$ because, as already suggested, the mass of the spectator is smaller than the mass of the inflaton, but in principle not negligible. In so doing, we obtain
\begin{equation}
\theta_{1\xi}(t)=\theta_I \frac{\Xi}{\xi} \!\left\{\!\frac{\left[\theta_1(t)\!-\!\alpha\theta_2(t)\right]\!+\!\Sigma\left[\theta_3(t)\!-\!\alpha\theta_4(t)\right]}{2 e^{\Gamma t/2}}\!\right\}\!,
\label{eqREDUCEDINFLATONIC}
\end{equation}
where we introduced the dimensionless parameter $\Xi=\xi\phi_I^2/\Omega^2$ and the following auxiliary terms
\begin{subequations}
\begin{align}
&\theta_1(t)=\cos(\Omega t)\sin^2(\alpha\Omega t)-2\frac{\Xi_\alpha(t)}{\Xi}(\Omega t)\sin(\Omega t),
\label{eqRENAME1}\\
&\theta_2(t)=\sin(\Omega t)\sin(2\alpha\Omega t),
\label{eqRENAME2}\\
&\theta_3(t)=\frac{2\sin(\Omega t)\sin^2(\alpha\Omega t)}{\alpha}+\cos(\Omega t)\sin(2\alpha\Omega t),
\label{eqRENAME3}\\
&\theta_4(t)=\sin(\Omega t)\cos(2\alpha\Omega t)\!-\!\frac{4\cos^2(\alpha\Omega t)\sin(2\Omega t)}{3e^{\Gamma t/2}},
\label{eqRENAME4}
\end{align}
\end{subequations}
and in Eq.~\eqref{eqRENAME1} we defined the dimensionless quantity
\begin{equation}
\Xi_{\alpha}(t)=\frac{\Xi}{2}\left[1+\frac{\sin(2\alpha\Omega t)}{2\alpha\Omega t}\right].
\label{eqEFFECTIVECOUPLING}
\end{equation}

Finally, labeling the background solution with
\begin{equation}
\theta_0(t)=\theta_Ie^{-\Gamma t/2}\cos(\Omega t),
\label{eqRENAME0}
\end{equation}

from Eqs.~\eqref{eqREDUCEDINFLATONIC}--\eqref{eqRENAME0}, we obtain the full inflatonic solution 
\begin{equation}
\theta(t)=\theta_0(t)+\xi\theta_{1\xi}(t).
\label{eqREDUCEDINFLATONIC2}
\end{equation}

In Eq.~\eqref{eqREDUCEDINFLATONIC2} (or better in Eq.~\eqref{eqREDUCEDINFLATONIC}) we can identify
\begin{itemize}
    \item[1)] a first-order correction $\propto\Xi\Sigma[\theta_3(t)-\alpha\theta_4(t)]$, due to the Yukawa-like coupling between $\phi$ and $R$, and
    \item[2)] a correction $\propto\Xi[\theta_1(t)-\alpha\theta_2(t)]$, due to the coupling between inflaton and spectator.
\end{itemize}
\begin{figure*}
\centering
\includegraphics[width=0.9\hsize,clip]{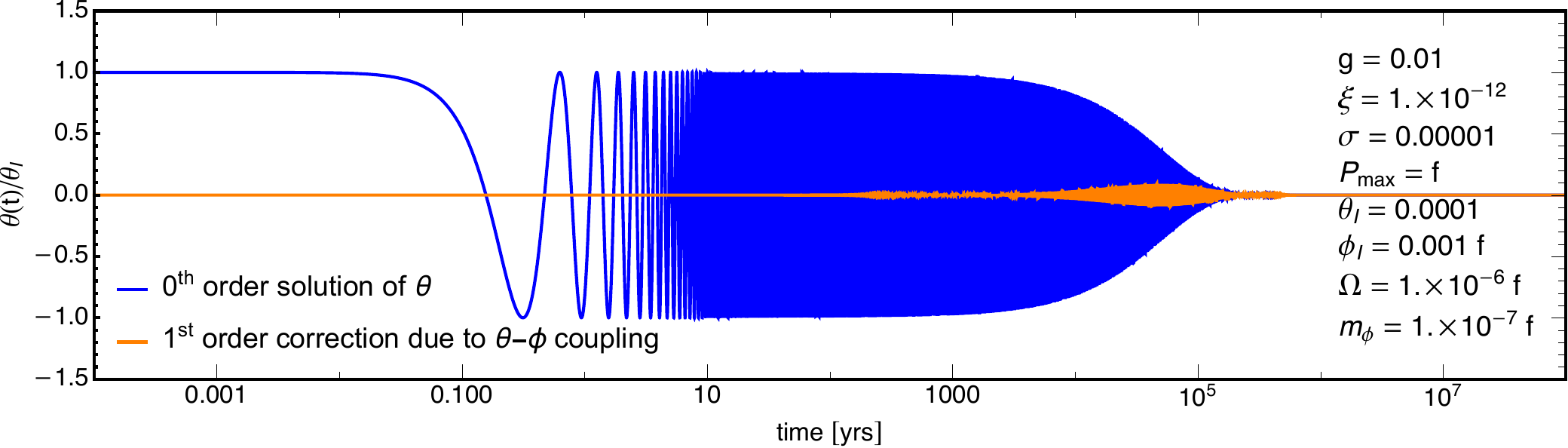}\\
\includegraphics[width=0.9\hsize,clip]{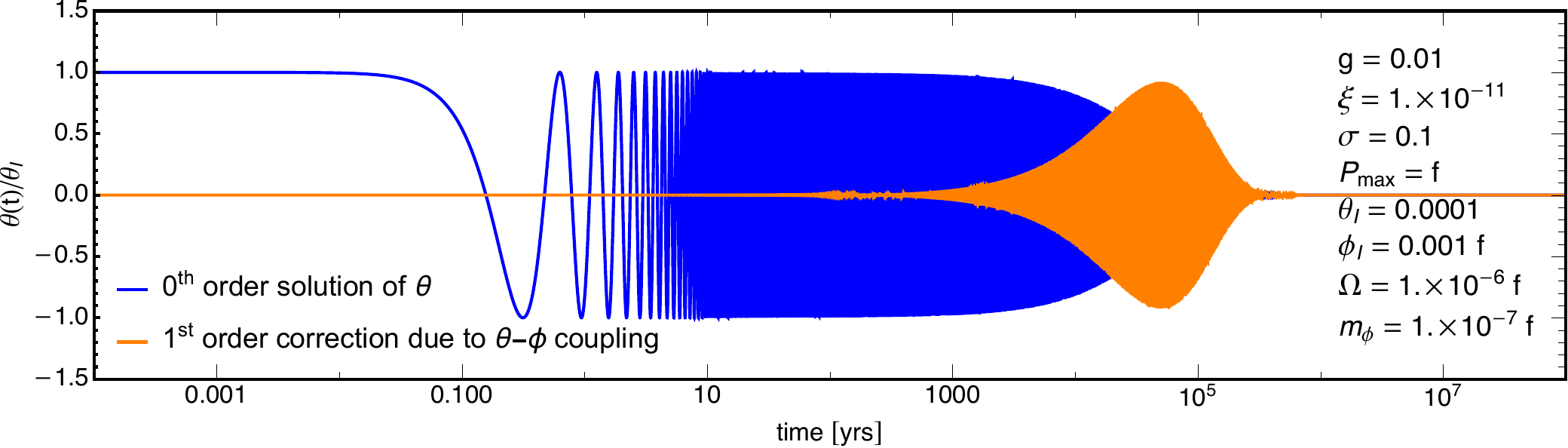}
\caption{Comparison between first-order correction for $\theta(t)$ due to the $\theta-\phi$ coupling and the background solution. \emph{Top}: for $g=0.01$, $\Omega=10^{-6}f$, $\xi=10^{-12}$, $\sigma=10^{-5}$, $\phi_I=10^{-3}f$ and $m_{\phi}=10^{-7}f$, the first-order solution is a perturbation of the background solution. \emph{Bottom}: for $g=0.01$, $\Omega=10^{-6}f$, $\xi=10^{-11}$, $\sigma=10^{-1}$, $\phi_I=10^{-3}f$ and $m_{\phi}=10^{-7}f$, the first-order correction is larger than the background solution, thus the perturbative method fails.}
\label{fig2}
\end{figure*}

We start with 2) and work in the regime of small $\alpha$. The first term in $\theta_1(t)$ and $\alpha\theta_2(t)$ scale as $\alpha^2$ and can be considered negligible. Conversely, the second term in $\theta_1(t)$ is of the order of the unity and, thus, represents the the leading correction in terms of baryon asymmetry. Moreover, this leading term contains $\Xi_{\alpha}(t)$ defined in Eq.~\eqref{eqEFFECTIVECOUPLING}, that can be viewed as time-dependent \emph{effective coupling constant} with the following limits:
\begin{itemize}
    \item[-] for $t\ll 1/m_{\phi}$, we have $\Xi_{\alpha}(t)\equiv\Xi$ and the inflaton has not enough time to oscillate and can be considered as a constant field, whereas
    \item[-] for $t\gg 1/m_{\phi}$, we have $\Xi_{\alpha}(t)\equiv\Xi/2$ and the oscillations of the inflaton become significant.
\end{itemize}
Equivalently, expliciting the dependency upon $\phi_I$ in Eq.~\eqref{eqEFFECTIVECOUPLING}, we can define the \emph{effective square value} of $\phi(t)$
\begin{equation}
\braket{\phi^2_0(t)}=\frac{\phi_I^2}{2}\left[1+\frac{\sin(2\alpha\Omega t)}{2\alpha\Omega t}\right],
\label{eqEFFECTIVECOUPLING2}
\end{equation}
which is equal to $\phi_I^2$ for an almost constant spectator field ($t\ll1/m_{\phi}$), whereas coinciding to the effective value, $\phi_I^2/2$, for a rapidly oscillating one ($t\gg 1/m_{\phi}$).

We will see that considering $\Xi$ rather than $\Xi_{\alpha}(t)$ does not produce any change in terms of baryon asymmetry production. Thus, one could be tempted to directly substitute $\Xi$ in lieu of $\Xi_{\alpha}(t)$. However, the dynamics of the inflaton seems to change if considering $\Xi_{\alpha}(t)$ or $\Xi$. In particular, as already suggested, the two in principle coincide only for a limited interval of time, $t\ll1/m_{\phi}$. However, in terms of produced baryon asymmetry, the interesting epoch is the one during which the inflaton mostly decays into fermion-antifermion pairs, that can be roughly estimated by the characteristic time $\tau=2/\Gamma$.
Therefore, we expect that if $\tau<1/m_{\phi}$ or $m_{\phi}<\Gamma$, then the approximation $\Xi_{\alpha}(t)\sim\Xi$ is valid for the whole duration of decays. However, as previously discussed, the most plausible scenario is when $m_{\phi}>\Gamma$, for which the expansion of the universe can be safaly neglected in the equation of motion for the spectator.
In conclusion, knowing whether $m_\phi<\Gamma$ or $m_\phi>\Gamma$ is not fundamental, because we are interested in the baryon asymmetry production and, in this respect, considering either $\Xi_{\alpha}(t)$ or $\Xi$ is equivalent.

\begin{figure*}
\centering
\includegraphics[width=0.9\hsize,clip]{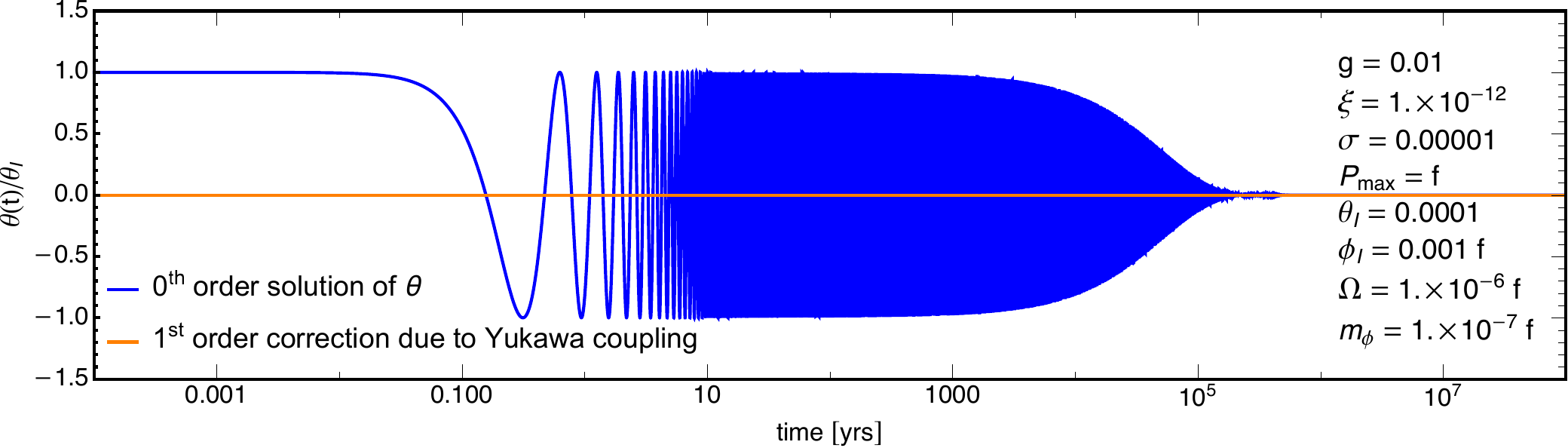}\\
\includegraphics[width=0.9\hsize,clip]{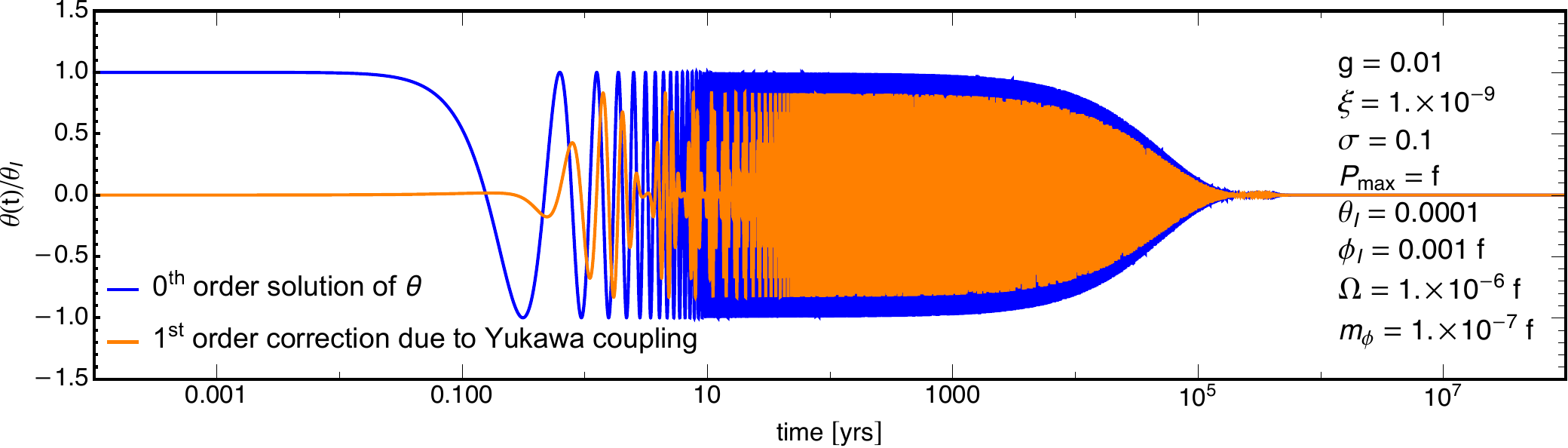}
\caption{Comparison between the first order correction for $\theta(t)$ due to the Yukawa-like $\phi$--$R$ coupling and the background solution. \emph{Top}: for $g=0.01$, $\Omega=10^{-6}f$, $\xi=10^{-12}$, $\sigma=10^{-5}$, $\phi_I=10^{-3}f$ and $m_{\phi}=10^{-7}f$, the first-order solution is much smaller than the background solution, thus the perturbative method succeeds. \emph{Bottom}: for $g=0.01$, $\Omega=10^{-6}f$, $\xi=10^{-9}$, $\sigma=10^{-1}$, $\phi_I=10^{-3}f$ and $m_{\phi}=10^{-7}f$, the other first-order correction becomes comparable to the background and the perturbative method fails.}
\label{fig3}
\end{figure*}

What really matters is to find the appropriate constraints on the model free parameters in order to ensure the validity of the perturbative regime. To this aim, we start with the first-order term  $\propto\Xi[\theta_1(t)-\alpha\theta_2(t)]$ of the inflatonic solution in Eqs.~\eqref{eqREDUCEDINFLATONIC} and \eqref{eqREDUCEDINFLATONIC2}.
As already discussed, the most relevant term for the baryon asymmetry is the second one of Eq.~\eqref{eqRENAME1}. Unlike the background solution -- oscillating with an amplitude that is exponentially decreasing with the cosmic time -- the amplitude of the second term of $\theta_1(t)$ follows a more complex behavior, increasing up to a maximum and then exponentially decaying. Denoting with $\mathcal{A}_0(t)=e^{-\Gamma t/2}$ the amplitude of the background solution and with $\mathcal{A}_1(t)=\Xi(\Omega t)e^{-\Gamma t/2}$ the amplitude of the corrective term\footnote{Notice that, here, we use $\Xi$ in lieu of $\Xi_{\alpha}(t)$, which licit when neglecting terms $\propto\alpha^2$ with respect to the leading one.}, at the characteristic time $\tau$, the latter is at its maximum $\mathcal{A}_1(\tau)=(2\Xi\Omega/\Gamma) \mathcal A_0(\tau)$. Therefore, to guarantee the validity of the perturbative regime we have to impose
\begin{equation}
\frac{2\Xi\Omega}{\Gamma}\ll1 \implies \xi\ll  \frac{\Omega\Gamma}{2\phi_I^2}=\frac{\Gamma}{f}\sim \frac{g^2\Omega}{f}=10^{-6}g^2.
\label{eqXI1}
\end{equation}
Fig.~\ref{fig2} compares the background solution with the above correction due to the coupling between inflaton and spectator fields. In particular, the top panel is obtained for a set of parameters satisfying Eq.~\eqref{eqXI1} for which the perturbative regime holds; conversely, the bottom panel is realized with a set of parameters not representative of Eq.~\eqref{eqXI1}, for which the perturbative method fails.

Now we consider the term $\propto\Xi\Sigma[\theta_3(t)-\alpha\theta_4(t)]$ in Eqs.~\eqref{eqREDUCEDINFLATONIC} and \eqref{eqREDUCEDINFLATONIC2}, due to the Yukawa-like coupling between $\phi$ and $R$.
This term is a small perturbation with respect to the background only if
\begin{equation}
\begin{split}
&\frac{\Xi\Sigma}{\alpha}\ll1\implies  \xi\sigma\ll 10^{-6}\left(\frac{m_{\phi}}{g \phi_I}\right)^2=\frac{1}{g^2}\left(\frac{m_{\phi}}{f}\right)^2.
\end{split}
\label{eqXI2}
\end{equation}
Finally, it is important to notice that Eqs.~\eqref{eqEOM16} and \eqref{eqXI1} must be valid too and, when combined, provide
\begin{equation}
\xi\sigma\ll 10^{-6}\frac{\Omega m_{\phi}}{\phi_I^2}=10^{-6}\frac{m_{\phi}}{f}\sim10^{-12}g^2\frac{m_{\phi}}{\Gamma}.
\label{eqXI3}
\end{equation}
In Fig.~\ref{fig3}, we can see that, when Eqs.~\eqref{eqXI2}--\eqref{eqXI3} hold, the first-order correction due to the Yukawa-like coupling is a small correction to the background solution (top panel); when the above constraints are not met, then the perturbative approach fails (bottom panel).

Once all the model parameters are well determined and satisfy the constraints dictated by the validity of perturbative approach, we proceed in computing explicitly the baryon asymmetry via the inflatonic solution in Eq.~\eqref{eqREDUCEDINFLATONIC2}. In particular we prove that the resulting baryon asymmetry coincides with that coming from using $\Xi$ in lieu of $\Xi_{\alpha}(t)$, in the limit of negligible terms $\propto\alpha^2$.

\section{Particle production}\label{sec:level5}

The baryon asymmetry can be computed as in the background model, but replacing the zero-order solution for $\theta(t)$ in Eq.~\eqref{eq5} with Eq.~\eqref{eqREDUCEDINFLATONIC2}.
In doing so, we use a semiclassical approach, in which: (a) $\theta$, $\phi$, and gravity are classical fields, and (b) the fermions $Q$ and $L$ are the only quantized fields.
Thus, within this framework, baryons and antibaryons can be produced only by, respectively, $Q-\overline{L}$ and $\overline{Q}-L$ (fermions and antifermions) pairs, produced by the decays of the inflaton.
Therefore, from Eq.~\eqref{eq6}, we can compute the average number densities of baryons $n(Q,\overline{L})$ and antibaryons $n(L,\overline{Q})$
\begin{subequations}
\begin{align}
\nonumber
&\!n(Q,\overline{L})=\frac{1}{V}\sum_{s_Q,s_L}\int\frac{d^3p_Q}{(2\pi)^32p_Q^0}\frac{d^3p_L}{(2\pi)^32p_L^0} \bigg{|}i\frac{gf}{\sqrt{2}}\times \\
&\!\hat{T}\!\left[\!\int \!d^4x\!\braket{Q(p_Q,\!s_Q\!),\overline{L}(p_L,\!s_L\!|\hat{\overline{Q}}(x)\hat{L}(x)e^{i\theta(x)}\!|0}\!\right]\!\bigg{|}^2\!,\!\label{eq34}\\
\nonumber
&\!n(L,\overline{Q})=\frac{1}{V}\sum_{s_Q,s_L}\int\frac{d^3p_Q}{(2\pi)^32p_Q^0}\frac{d^3p_L}{(2\pi)^32p_L^0}\bigg{|}i\frac{gf}{\sqrt{2}}\times \\
&\!\hat{T}\!\left[\!\int \!d^4x\!\braket{L(p_L,\!s_L\!)\!,\!\overline{Q}(p_Q,\!s_Q\!)|\hat{\overline{L}}(x)\hat{Q}(x)e^{-i\theta(x)\!}|0}\!\right]\!\bigg{|}^2\!,\!
\label{eq35}
\end{align}
\end{subequations}
where $\hat T$ is the time-ordering operator.

Since $g\ll1$, the masses of the fermions, $(m_Q,m_L)\sim gf$, are negligible compared to the inflaton mass $\Omega$.
Hence, Eqs.~\eqref{eq34}--\eqref{eq35} can be evaluated in the massless limit. In so doing, by using the standard canonical quantization for the fermionic fields $Q$ and $L$, we obtain
\begin{subequations}
\begin{align}
&n(Q,\overline{L})=\frac{g^2f^2}{2\pi^2}\int \omega^2\bigg{|}\int_{-\infty}^{+\infty}dte^{i\theta}e^{2i\omega t}\bigg{|}^2d\omega,
\label{eq36}\\
&n(L,\overline{Q})=\frac{g^2f^2}{2\pi^2}\int \omega^2\bigg{|}\int_{-\infty}^{+\infty}dte^{-i\theta}e^{2i\omega t}\bigg{|}^2d\omega.
\label{eq37}
\end{align}
\end{subequations}
Consequently, the baryon asymmetry results to be
\begin{equation}
n_B=n(Q,\overline{L})-n(L,\overline{Q})=\frac{g^2f^2}{2\pi^2}\int_0^{+\infty}\mathcal{F}(\omega)\omega^2d\omega,
\label{eq38}
\end{equation}
with $\mathcal{F}(\omega)$ defined as
\begin{equation}
\mathcal{F}(\omega)\!=\!\int_{-\infty}^{+\infty}\!dt \,dt^\prime e^{2i\omega(t-t')}\!\left[ e^{i\Delta\theta(t,t^\prime)}\!-\!e^{-i\Delta\theta(t,t^\prime)} \right]\!.
\label{eq39}
\end{equation}
where both $t$ and $t^{'}$ time-variables span within $\pm\infty$ and $\Delta\theta(t,t^\prime) = \theta(t)-\theta(t')$.
Since during reheating epoch the inflaton field makes small oscillation around the minimum of its potential $\theta=0$, we can expand $\theta(t)$ in $\mathcal{F}(\omega)$ up to second order around $\theta=0$. In this way, we get
\begin{equation}
\mathcal{F}(\omega)=2\text{Re}\bigg{\{}\frac{\tilde{\theta}(2\omega)\tilde{\theta}^{2*}(2\omega)}{i}\bigg{\}},
\label{eq40}
\end{equation}
where $\tilde{\theta}^n(2\omega)$, with $n=\{1,2\}$, is given by
\begin{equation}
\tilde{\theta}^n(2\omega)=\int_{-\infty}^{+\infty}dt e^{2i\omega t}\theta^n(t).
\label{eq41}
\end{equation}
Let $\tilde{\theta}_0(2\omega)$ and $\tilde{\theta}^{2*}_0(2\omega)$ be respectively the zero-orders of Eq.~\eqref{eq41}, whose expressions can be found in Ref. \cite{Dolgov:1996qq}, and let $\mathcal{F}_0(\omega)$ be the expression for $\mathcal{F}(\omega)$ obtained for $\tilde{\theta}_0(2\omega)$ and $\tilde{\theta}_0^{2*}(2\omega)$.
Then, the baryon asymmetry with $\mathcal{F}(\omega)=\mathcal{F}_0(\omega)$ leads to the result in Eq.~\eqref{eq7}. The first-order corrections, appearing when using Eq.~\eqref{eqREDUCEDINFLATONIC2} for $\theta(t)$, can be found by computing the corresponding expressions for $\tilde{\theta}(2\omega)$ and $\tilde{\theta}^{2*}(2\omega)$. In particular, for $\tilde{\theta}(2\omega)$, we have
\begin{equation}
\begin{split}
&\tilde{\theta}(2\omega)=\tilde{\theta}_0(2\omega)+\Xi\big\{[\tilde{\theta}_1(2\omega)-\alpha\tilde{\theta}_2(2\omega)]+\\
&+\Sigma[\tilde{\theta}_3(2\omega)-\alpha\tilde{\theta}_4(2\omega)]\big\},
\end{split}
\label{eq43}
\end{equation}
where the functions $\tilde{\theta}_j(2\omega)$, with $j=\{1,2,3,4\}$, are defined from Eqs.~\eqref{eqRENAME1}--\eqref{eqRENAME4}
\begin{equation}
\tilde{\theta}_j(2\omega)=\frac{\theta_I}{2}\int_0^{+\infty}e^{2i\omega t}e^{-\Gamma t/2}\theta_j(t)dt.
\label{eqDEF1}
\end{equation}
To simplify the notation, we define $\gamma=\Gamma/\Omega$, $x=\omega/\Omega$, and $y=2ix-\gamma/2$. Thus, the term $\tilde{\theta}_j(2\omega)$ read
\begin{subequations}
\begin{align}
\nonumber
&\tilde{\theta}_1(2\omega)=\frac{\theta_I}{\Omega}\!\left[\frac{y/8}{(1+2\alpha)^2+y^2}\!+\!
\frac{y/8}{(1-2\alpha)^2+y^2}\!-\!\frac{y/4}{1+y^2}+\right.\\
&\left.\frac{y}{(1+y^2)^2}+\frac{y}{[(1+2\alpha)^2+y^2][(1-2\alpha)^2+y^2]}\right],
\label{eqTILDETHETA1}\\
&\tilde{\theta}_2(2\omega)=-\frac{2\theta_I\alpha y}{\Omega[y^2+(1+2\alpha)^2][y^2+(1-2\alpha)^2]},
\label{eqTILDETHETA2}\\
\nonumber
&\tilde{\theta}_3(2\omega)=\frac{\theta_I}{4\alpha\Omega}\left[\frac{2}{1+y^2}-\frac{(1-\alpha)(1+2\alpha)}{y^2+(1+2\alpha)^2}\right.\\
&\left.-\frac{(1+\alpha)(1-2\alpha)}{y^2+(1-2\alpha)^2}\right],
\label{eqTILDETHETA3}\\
\nonumber
&\tilde{\theta}_4(2\omega)=\frac{\theta_I}{\Omega}\left[\frac{(1+2\alpha)/4}{y^2+(1+2\alpha)^2}+\frac{(1-2\alpha)/4}{y^2+(1-2\alpha)^2}-\frac{2/3}{y^2+4}\right.\\
&\left. +\frac{(1+\alpha)/3}{y^2+4(1+\alpha)^2}+\frac{(1-\alpha)/3}{y^2+4(1-\alpha)^2}\right].
\label{eqTILDETHETA4}
\end{align}
\end{subequations}

We start focusing on the terms of Eq.~\eqref{eq43} proportional to $\Xi$ in the inflatonic solution.
\begin{itemize}
    \item[-] The first line of Eq.~\eqref{eqTILDETHETA1}, including its derivative with respect to $\alpha$, vanishes for $\alpha=0$, therefore, it scales as $\alpha^2$. Conversely, the two terms in the second line of Eq.~\eqref{eqTILDETHETA1} do not vanish for $\alpha=0$ and contributes at order zero in $\alpha$, respectively due to the constant and the time-dependent part of $\Xi_{\alpha}(t)$.
    \item[-] Eq.~\eqref{eqTILDETHETA2} scales as $\alpha$ and  in Eq.~\eqref{eq43} is also multiplied by $\alpha$, thus it scales as $\propto\alpha^2$ and is negligible with respect to the leading contribution of $\tilde{\theta}_1(2\omega)$.
    \item[-] Therefore, the leading contribution to the baryon asymmetry is at zero order in $\alpha$ and comes from the second line of $\tilde{\theta}_1(2\omega)$ in Eq.~\eqref{eqTILDETHETA1}.
\end{itemize}

About the terms of Eq.~\eqref{eq43} due to the Yukawa-like coupling, the situation is slightly different.
\begin{itemize}
    \item[-] The part of Eq.~\eqref{eqTILDETHETA3} within square brackets, including its derivative with respect to $\alpha$, vanishes for $\alpha=0$. Therefore, it scales as $\alpha^2$ and, with the factor $\alpha^{-1}$ in front, the whole $\tilde{\theta}_3(2\omega)$ scales as $\alpha$.
    \item[-] Imposing $\alpha=0$ in Eq.~\eqref{eqTILDETHETA4} provides $\tilde{\theta}_4(2\omega)$ at zero order in $\alpha$, but it appears in Eq. \ref{eq43} multiplied by $\alpha$, contributing finally at order $\alpha$, like $\tilde{\theta}_3(2\omega)$.
    \item[-] Therefore, both $\tilde{\theta}_3(2\omega)$ and $\alpha\tilde{\theta}_4(2\omega)$ contribute to the baryon asymmetry at order $\alpha$.
\end{itemize}

Similar considerations can be deduced by computing $\tilde{\theta}^{2*}(2\omega)$. In particular, it can be written as
\begin{equation}
\begin{split}
&\tilde{\theta}^{2*}(2\omega)=\tilde{\theta}_0^{2*}(2\omega)+2\Xi\Big\{\left[\tilde{\theta}_{01}^{2*}(2\omega)-\alpha\tilde{\theta}_{02}^{2*}(2\omega)\right]+\\
&+\Sigma\left[\tilde{\theta}_{03}^{2*}(2\omega)-\alpha\tilde{\theta}_{04}^{2*}(2\omega)\right]\Big\},
\end{split}
\label{eq44}
\end{equation}
where the functions $\tilde{\theta}_{0j}^{2*}(2\omega)$, with $j=1,2,3,4$, are
\begin{equation}
\tilde{\theta}^{2*}_{0j}(2\omega)=\frac{\theta_I}{2}\int_0^{+\infty}e^{-2i\omega t}e^{-\Gamma t/2}\theta_j(t)\theta_0(t)dt.
\label{eqDEF2}
\end{equation}
Defining $y_0=2ix+\gamma$ and $y_1=2ix+3\gamma/2$, we obtain
\begin{subequations}
\begin{align}
\nonumber
&\tilde{\theta}_{01}^{2*}(2\omega)=\frac{\theta_I^2}{\Omega}\bigg{\{} \frac{1}{8y_0}-\frac{y_0/8}{y_0^2+4\alpha^2}+\frac{y_0/8}{y_0^2+4}+\\
\nonumber
&\left.-\frac{y_0/16}{y_0^2+4(1+\alpha)^2}-\frac{y_0/16}{y_0^2+4(1-\alpha)^2}+\right.\\
&-\frac{y_0}{(y_0^2+4)^2}\!-\!\frac{y_0}{[y_0^2+4(1+\alpha)^2][y_0^2+4(1-\alpha)^2]}\!\bigg{\}},
\label{eqTILDETHETA01}\\
&\tilde{\theta}_{02}^{2*}(2\omega)=\frac{2\theta_I^2\alpha}{\Omega}\frac{y_0}{[y_0^2+4(1+\alpha)^2][y_0^2+4(1-\alpha)^2]},
\label{eqTILDETHETA02}\\
\nonumber
&\tilde{\theta}_{03}^{2*}(2\omega)=\frac{\theta_I^2}{2\alpha\Omega}\bigg{\{}\frac{\alpha^2}{y_0^2+4\alpha^2}+\frac{1}{y_0^2+4}+\\
&-\frac{1}{2}\left[\frac{1-\alpha^2}{y_0^2+4(1+\alpha)^2}+\frac{1-\alpha^2}{y_0^2+4(1-\alpha)^2}\right]\bigg{\}},
\label{eqTILDETHETA03}\\
\nonumber
&\tilde{\theta}_{04}^{2*}(2\omega)=\frac{\theta_I^2}{4\Omega}\bigg{[}\frac{1+\alpha}{y_0^2+4(1+\alpha)^2}+\frac{1-\alpha}{y_0^2+4(1-\alpha)^2}\bigg{]}+\\
\nonumber
&-\frac{\theta_I^2}{6\Omega}\left[\frac{3}{y_1^2+9}+\frac{3/2+\alpha}{y_1^2+(3+2\alpha)^2}+\frac{3/2-\alpha}{y_1^2+(3-2\alpha)^2}+\right.\\
&\left.\frac{1}{y_1^2+1}+\frac{1/2+\alpha}{y_1^2+(1+2\alpha)^2}+\frac{1/2-\alpha}{y_1^2+(1-2\alpha)^2}\right].
\label{eqTILDETHETA04}
\end{align}
\end{subequations}

Once again, we check that the leading contributions to the baryon asymmetry correction.
\begin{itemize}
    \item[-] Among the ones proportional to $\Xi$, the leading term is at zero order in $\alpha$ and comes from $\tilde{\theta}_{01}^{2*}(2\omega)$, specifically from the terms in the last line of Eq.~\eqref{eqTILDETHETA01}, containing again $\Xi_{\alpha}(t)$.
    \item[-] Regarding the Yukawa-like coupling terms, they all provide a contribution of the order $\alpha$.
\end{itemize}

Having determined the expressions for $\tilde{\theta}(2\omega)$ and $\tilde{\theta}^{2*}(2\omega)$, we evaluate the full expression for $\mathcal{F}(\omega)$ via Eq.~\eqref{eq40}.
In particular, the product between $\tilde{\theta}(2\omega)$ and $\tilde{\theta}^{2*}(2\omega)$ at the first order in $\Xi$ is given by
\begin{align}
\nonumber
&\tilde{\theta}(2\omega)\tilde{\theta}^{2*}(2\omega)=\tilde{\theta}_0(2\omega)\tilde{\theta}_0^{2*}(2\omega)+\Xi\big{\{}\tilde{\theta}_0^{2*}(2\omega)\times\\
\nonumber
&\times\big{[}\big{(}\tilde{\theta}_1(2\omega)-\alpha\tilde{\theta}_2(2\omega)\big{)}+\Sigma\big{(}\tilde{\theta}_3(2\omega)-\alpha\tilde{\theta}_4(2\omega)\big{)}\big{]}+\\
\nonumber
&+2\tilde{\theta}_0(2\omega)\big{[}\big{(}\tilde{\theta}_{01}^{2*}(2\omega)-\alpha\tilde{\theta}_{02}^{2*}(2\omega)\big{)}+\\
&+\Sigma\big{(}\tilde{\theta}_{03}^{2*}(2\omega)-\alpha\tilde{\theta}_{04}^{2*}(2\omega)\big{)}\big{]}\big{\}},
\label{eqPRODUCTTHETATILDE}
\end{align}
that at zero order reproduces the background result for the baryon asymmetry, whereas the other contributions are responsible for the first-order correction.

The baryon asymmetry can be now computed via Eqs.~\eqref{eqTILDETHETA1}--\eqref{eqTILDETHETA4}, \eqref{eqTILDETHETA01}--\eqref{eqTILDETHETA04}, and \eqref{eqPRODUCTTHETATILDE}. In view of the above considerations, we expect the baryon asymmetry to contain a first-order correction proportional to $\Xi$ and another one proportional to $\alpha\Xi\Sigma$ with respect to the result obtained in the background model.

Working with the limit $g\ll1$ (or $\Omega\gg\Gamma$) and neglecting the order $\alpha^2$, the total baryon asymmetry reads
\begin{equation}
n_B = n_B^{(0)} \left[1+\Xi\left(1+\frac{8\pi\alpha\Sigma}{3g^2}\right)\right],
\label{eq45}
\end{equation}
where $n_B^{(0)}=\Omega g^2f^2\theta_I^3/(16\pi)$ refers to the expression in Eq.~\eqref{eq7}, representing the background model. Full calculations can be found in Appendix \ref{app3}.

As anticipated, the first-order correction is composed of a term proportional to $\Xi$ and another one proportional to $\alpha\Sigma\Xi$.
To evaluate the order of magnitude of the correction, the model parameters shall conform to the adopted perturbative approach regime, that can be resumed with the following constraints
\begin{equation}
\begin{cases}
\Sigma\ll1\\
\Xi\ll g^2\\
\Xi\Sigma\ll\alpha
\end{cases}\implies
\begin{cases}
\sigma\ll (m_{\phi}/f)/g^2 \\ \xi\ll10^{-6}g^2 \\ \xi\sigma \ll (m_{\phi}/f)^2/g^2
\end{cases}.
\label{eqCONSTRAINTS}
\end{equation}

As expected, when neglecting the expansion of the universe, the baryon asymmetry production is not significantly altered with respect to the background by the presence of a complex scalar spectator field Yukawa-like coupled to the gravitational field and weakly interacting with the inflaton.
In particular, to evaluate whether the contribution of the Yukawa-like coupling is significant or not, it is sufficient to assess if $\alpha\Sigma/g^2\sim1$. Provided that we consider the case $m_{\phi}>\Gamma$, that is roughly the same to take $\alpha>g^2$, hence Eq.~\eqref{eqEOM16} does not exclude the possibility that $\Sigma$ would become greater than $g^2/\alpha$. In particular, it is not difficult to prove that $\alpha\Sigma/g^2\sim 10^6\sigma$, becoming relevant for $\sigma\ge10^{-6}$ and thus possibly leading to a significant first-order correction to the baryon asymmetry from the Yukawa-like coupling.
Nevertheless, the whole first-order correction in Eq.~\eqref{eq45} might not be comparable to the unity.

Before of including the effect of the expansion of the universe, we can draw further considerations about the first-order correction proportional only to $\Xi$. As already discussed, the first-order correction to the inflatonic solution contains the \emph{effective coupling constant} $\Xi_{\alpha}(t)$ defined in Eq.~\eqref{eqEFFECTIVECOUPLING}, where the contributions to the baryon asymmetry from the constant and the time-dependent terms are comparable, at the leading order. This can be also regarded as considering $\Xi$ in lieu of $\Xi_{\alpha}(t)$ or, equivalently, $\braket{\phi_0^2(t)}=\phi_I^2$.
With this in mind, the first-order correction proportional to only $\Xi$ can be regarded as due to a modification to the mass of the inflaton. This can be easily understood by looking at the equation of motion for the inflaton in Eq.~\eqref{eqEOMs}. There, the mass of the inflaton, $\Omega$, results corrected by a factor $\sqrt{1+2\xi\phi^2/\Omega^2}$ due to the presence of the spectator field.
Moreover, by replacing $\braket{\phi_0^2(t)}=\phi_I^2$, the mass of the inflaton results corrected by a constant factor $\sqrt{1+2\xi\phi_I^2/\Omega^2}\approx 1+\Xi$, where we used the definition of $\Xi$ and the constrain $\Xi\ll1$. This term is exactly equal to the correction to the baryon asymmetry that we obtain, of course without considering the Yukawa-like coupling.

\section{The mass mixing and consequences of cosmic expansion}\label{sec:level6}

We hereby discuss the effects of the mass-mixing, consequently restoring the universe expansion in our framework. The need of mass-mixing turns out to be relevant as it uniquely affects the quantized fields of our model.

Nevertheless, since the fermions $Q$ and $L$ are not mass eigenstates in the Lagrangian, after that the inflaton decays in fermion-antifermion pairs, a fermion produced as $Q$ may rotate in $L$ and \emph{vice versa}.
So, if the fields $Q$ and $L$ do not immediately decay into stable lighter mass particles with appropriate quark quantum numbers, they may have the chance to mix with each other.
Thus, we need to find the mass eigenstates and, then, compute once more the corresponding baryon asymmetry.

The general procedure is not altered by the introduction of the spectator field, if compared with the output of the background model \cite{Dolgov:1996qq} and, therefore, the corrective factor due to the mixing of baryon asymmetry mass is $[(1-\epsilon^2)/(1+\epsilon^2)]^2$, as reported in Sect.~\ref{sec:level2}.

In so doing, Eq.~\eqref{eq45} becomes
\begin{equation}
n_B=n_B^{(0)}\bigg{[}1+\Xi\bigg{(}1+\frac{8\pi\alpha\Sigma}{3g^2}\bigg{)}\bigg{]}\bigg{(}\frac{1-\epsilon^2}{1+\epsilon^2}\bigg{)}^2.
\label{eq46}
\end{equation}
The evaluate the above asymmetry, we only need to explicitly compute $\theta_I$.
In the regime in which the expansion of the universe is negligible with respect to the decay rate of the inflaton, i.e., $H\le \Gamma$, the initial value $\theta_I$ can be evaluated imposing that the baryon asymmetry  starts when $H=\Gamma$. This leads to $\theta_I=\sqrt{3/(4\pi)}\Gamma M_{\text{Pl}}/(f\Omega)$.

Observational constraints on the baryon asymmetry are always given as normalized to the entropy density of the universe $s=2\pi^2g_*T_{\text{RH}}^3/45$ computed at the reheating epoch temperature $T_{\text{RH}}=0.2(200/g_*)^{1/4}\sqrt{M_{\text{Pl}}\Gamma}$, with $g^\star\approx100$ being the effective numbers of relativistic species in thermal equilibrium with photons.

Consequently, using Eq.~\eqref{eq46} with the definition of $s$, the normalized baryonic asymmetry $\eta=n_B/s$ reads
\begin{equation}
\eta\simeq 3\times10^{-5}\frac{g^5 M_{\text{Pl}}^{3/2}}{\Omega^{1/2} f}\!\bigg{[}1+\Xi\bigg{(}\!1+\!\frac{8\pi\alpha\Sigma}{3g^2}\!\bigg{)}\bigg{]}\biggl(\!\frac{1-\epsilon^2}{1+\epsilon^2}\!\biggl)^2.
\label{eq48}
\end{equation}

We now include the expansion of the universe. Up to now, we have assumed particle production starting for $H=\Gamma$. However, it should have started much earlier, when $H=\Omega$, namely when the inflaton field started to oscillate around the minimum of its potential, which means that $\theta_I$ is modified by a factor $\Omega/\Gamma=8\pi/g^2$ and, consequently, $\eta$ by a factor $(8\pi/g^2)^3$.
Conversely, the value of $\phi_I$ is not altered because the spectator is just oscillating, albeit stable.

Assuming that the inflaton decay becomes significant at $H = \Omega$, the ensuing cosmic expansion leads to a progressive dilution of the baryon asymmetry generated at the time of maximal asymmetry production, $t_a \sim \Omega^{-1}$, up to the epoch of maximal entropy release, $t_b \sim \Gamma^{-1}$.
The dilution factor is $(a_a/a_b)^3\sim(t_a/t_b)^2\sim(\Gamma/\Omega)^2$ \cite{Dolgov:1996qq}, since the universe at reheating is assumed to be matter dominated with $a\propto t^{2/3}$.
Therefore, Eq.~\eqref{eq48} is shifted by a total factor $(\Omega/\Gamma)^3(\Gamma/\Omega)^2=\Omega/\Gamma=8\pi/g^2$, similarly to the background model. For the sake of completeness, we here remak a crucial difference though, namely $\theta_I$ is also contained in the definition of $\Sigma$, therefore,  $\Omega/\Gamma=8\pi/g^2$ also appears in front of $\Sigma$, thus yielding
\begin{equation}
\eta\simeq 10^{-3}\frac{g^3 M_{\text{Pl}}^{3/2}}{\Omega^{1/2} f}\! \bigg{[}\!1+\Xi\bigg{(}\!1\!+\!\frac{64\pi^2\alpha\Sigma}{3g^4}\!\bigg{)}\!\bigg{]}\!\biggl(\!\frac{1-\epsilon^2}{1+\epsilon^2}\!\biggl)^2.
\label{eq66}
\end{equation}

To establish if, by considering the expansion of the universe, the first-order correction becomes relevant, we consider once again the constraints in Eq.~\eqref{eqCONSTRAINTS}.
In particular, the first-order correction due to the Yukawa-like coupling might become relevant\footnote{The first-order correction not due to the Yukawa-like coupling is much smaller than the unity since $\Xi\ll1$.} for $g^2/(\alpha\Sigma)$ comparable to $\Xi/g^2$, that \emph{a priori} is allowed by Eq.~\eqref{eqCONSTRAINTS}. If this is fulfilled, we thus have
\begin{equation}
\begin{split}
&\frac{\Xi}{g^2}\ge\frac{g^2}{\alpha\Sigma}\implies
\sigma\xi\ge10^{-6}\frac{\Omega\Gamma}{\phi_I^2}\sim 10^{-12}g^2.
\end{split}
\label{eqFINAL}
\end{equation}

\begin{figure*}
\centering
\includegraphics[width=0.85\hsize,clip]{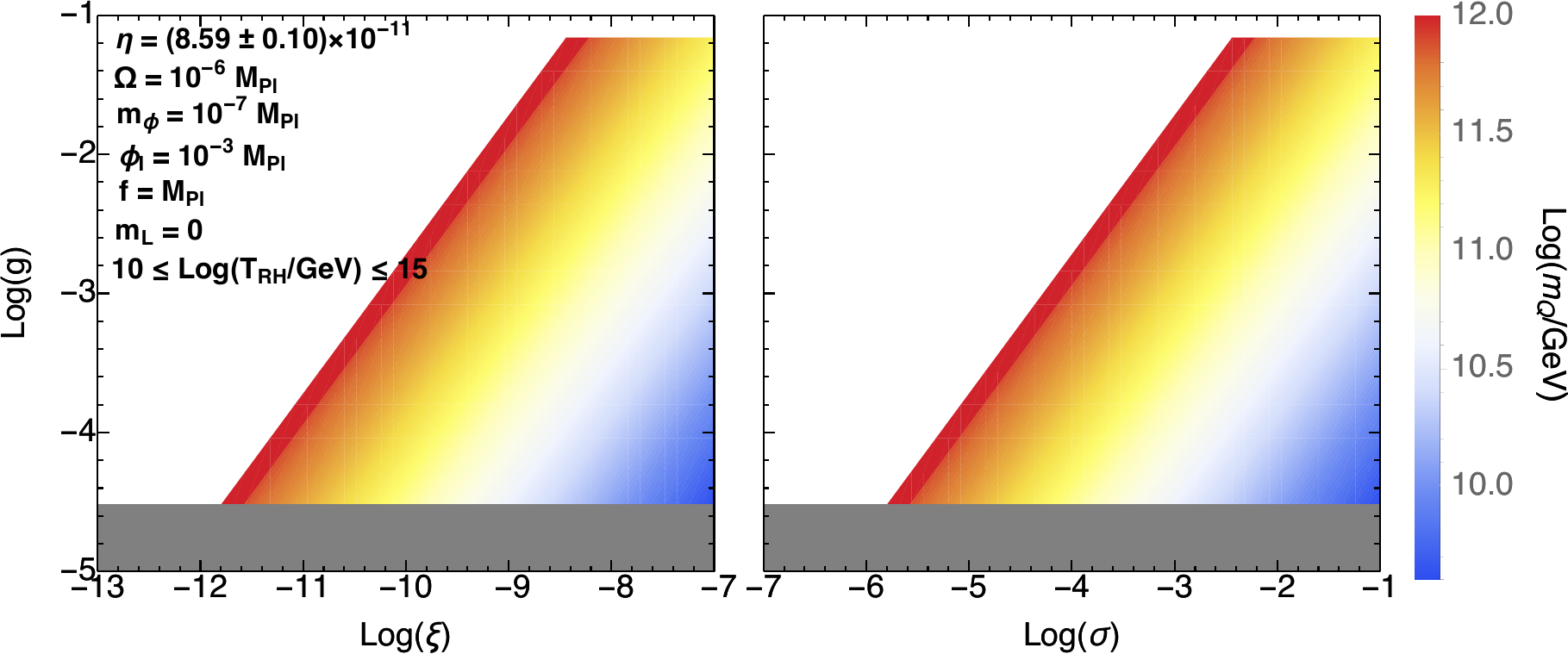}\\
\caption{Contour plot of the allowed region in the $\log{(g)}-\log{(\xi)}$ and $\log{(g)}-\log{(\sigma)}$ planes, compatible with the observational constraint $\eta=(8.59\pm0.10)\times 10^{-11}$. The chosen parameters are summarized in the legend. The gray areas are excluded by the lower bound on the reheating temperature $T_{\rm RH}=10^{10}$~GeV.}
\label{fig4}
\end{figure*}

The most stringent constraints on the product between $\xi$ and $\sigma$ are given by Eqs.~\eqref{eqXI2}--\eqref{eqXI3}. Among the two, it is not difficult to see that Eq.~\eqref{eqXI3} represents a stronger constraint with respect to Eq.~\ref{eqXI2} for $\alpha> g^2$ and \emph{vice versa}. As previously argued, the most plausible scenario is that $\alpha > g^2$, since the coupling $g$ takes typically very small values, while $\alpha$ can in principle be of order smaller or close than unity. For this reason, throughout our computations we neglect terms $\sim\alpha^2$.

On the contrary, consistently with the underlying background dynamics, we systematically neglect terms proportional to $\gamma = \Gamma/\Omega$ with respect to unity. As a consequence, Eq.~\eqref{eqXI3} is expected to impose a more stringent constraint than Eq.~\eqref{eqXI2}. Further, the condition $\alpha > g^2$ directly implies $m_{\phi} > \Gamma$, playing a central role to make the model consistent. In particular, this hierarchy allows Eq.~\eqref{eqFINAL} to hold without conflicting with the bound imposed by Eq.~\eqref{eqXI3}.

\begin{figure*}
\centering
\includegraphics[width=0.7\hsize,clip]{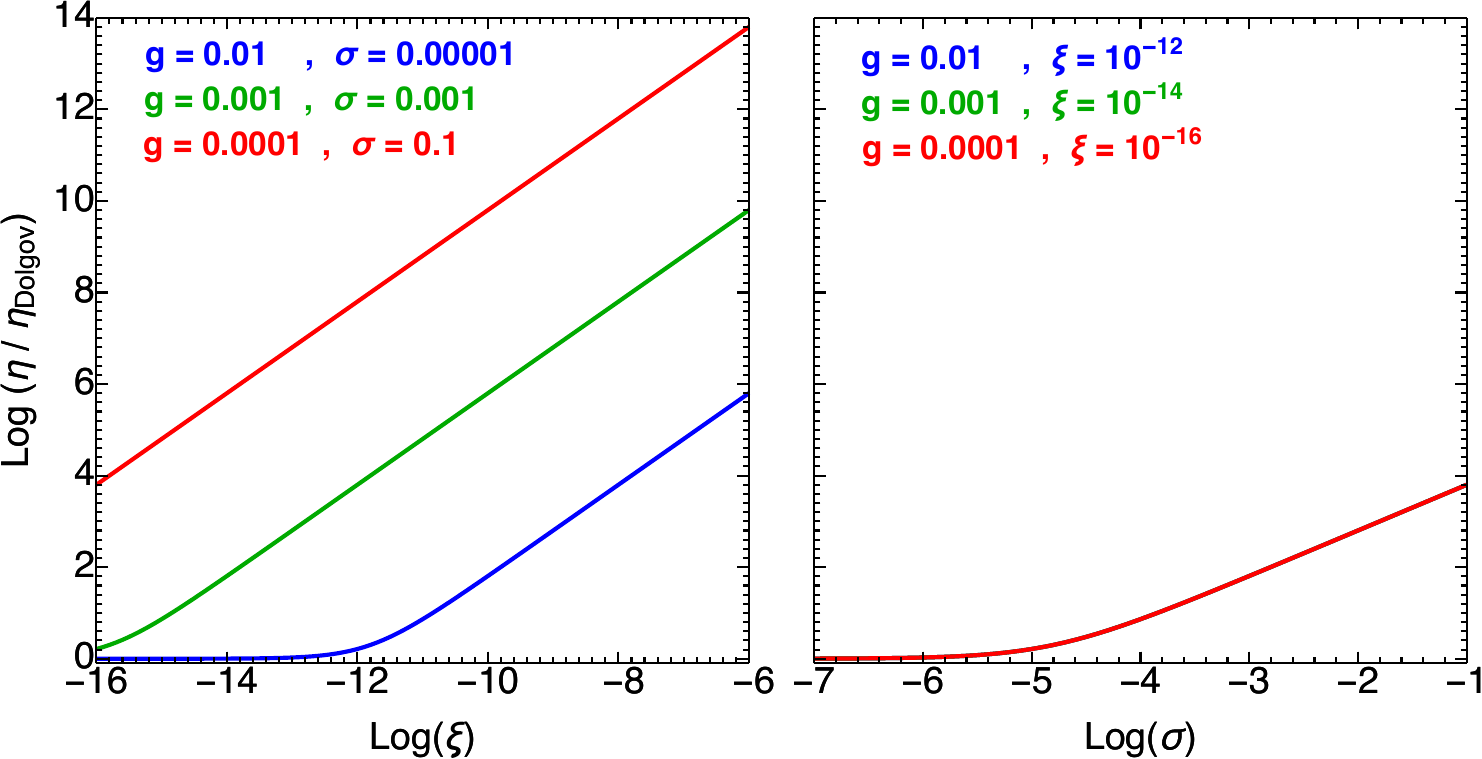}
\caption{Ratio between $\eta$ from Eq.~\eqref{eq66} and the background result \cite{Dolgov:1996qq} as function of $\xi$ and $\sigma$, for fixed values $g=0.01$, $g=0.001$, $g=0.0001$, and $\phi_I=10^{-3}f$. In the two plots, left and right, the values of $\sigma$ and $\xi$ respectively are fixed in agreement with the constraints imposed by the perturbative regime for each value of $g$.}
\label{fig5}
\end{figure*}

In Fig.~\ref{fig4}, we show contour plots displaying the allowed values of the free parameters $g$, $\sigma$ and $\xi$ such that the baryon asymmetry from Eq.~\eqref{eq66} agrees with the experimental value. Moreover, the fermionic mass $m_Q$ is ensured free to vary too, whereas we set $m_L=0$.

In Fig.~\ref{fig5}, we show the baryon asymmetry from Eq.~\eqref{eq66} normalized to the background for (left) selected values of $g$ and $\sigma$ and $\xi$ free to vary, and (right) selected values of $g$ and $\xi$ and $\sigma$ free to vary.
All the parameters used in the plots definitely respect the constraints imposed by the perturbative regime. Therefore, by virtue of the constraint in Eq.~\eqref{eqXI1}, the left plot of Fig.~\ref{fig5} implies that for decreasing values of the coupling constant $g=\{10^{-2},10^{-3},10^{-4}\}$ -- corresponding to $\xi\ll\{10^{-10},10^{-12},10^{-14}\}$ -- the baryon asymmetry is $\sim\{1,3,5\}$ orders of magnitude larger than background model, respectively. This can be also understood analytically by considering the Yukawa-like coupling correction to be the dominant one. Rewriting the first constraint of Eq.~\eqref{eqCONSTRAINTS} as $\Xi\sim 10^{-2}g^2$ or $\sigma\sim 10^{-9}/g^2$, with $m_{\phi}/f=10^{-7}$, from Eq.~\eqref{eq66} we obtain
\begin{equation}
\frac{\eta}{\eta_{\text{Dolgov}}}\sim 1+\frac{\Xi}{g^2}\frac{\alpha\Sigma}{g^2}\frac{64\pi^2}{3}\sim 1+\frac{10^{-5}}{g^2},
\label{eqFIRSTESTIMATE}
\end{equation}
from which the baryon asymmetry increases with respect to the background, even considering parameters allowed in the perturbative regime, for $g^2\le 10^{-5}$, as clearly shown in Fig.~\ref{fig5} for decreasing values of $g$.

Nevertheless, Fig.~\ref{fig4} suggests that the region of the parameter space compatible with the experimental baryon asymmetry does not exactly match the boundaries delimited by the perturbative regime.
On the one hand, this is a consequence, specially for $g$, of recent bounds on the reheating temperature, i.e. $10\leq \log(T_{\rm RH}/{\rm GeV})\leq15$ \cite{DiMarco:2020hru}. In fact, the excluded gray areas in Fig.~\ref{fig4} are determined by excluding $\log(T_{\rm RH}/{\rm GeV})\leq10$, thus relaxing this lower bound may provide an overlap with the constraints from the perturbative regime.
On the other hand, this means that, even improving also significantly the baryogenesis, introducing a scalar spectator field seems to be not enough in order to well reproduce the experimental data.
In so doing, we are just able to improve of some orders of magnitude the result for the baryon asymmetry obtained in the background model. In particular, we can see that by decreasing the coupling constant $g$, the correction results greater. This is an interesting and important fact, because very small values of the coupling constant $g$ are more plausible with respect to higher ones, in agreement with Planck observations \cite{Planck:2018jri}. 

\section{Final remarks}\label{sec:level7}

In this work, we investigated the framework of spontaneous baryogenesis by extending the background with a subdominant spectator field, under the form of a complex scalar weakly coupled to the inflaton and non-minimally coupled to gravity. Interactions between the spectator and standard model fermions were neglected under the assumption that they are highly suppressed and phenomenologically irrelevant at the scales considered.

Specifically, we introduced a coupling of the form $\theta^2 |\phi|^2$ between the inflaton and the spectator field, which induces a correction to the effective inflaton mass in its equation of motion. This choice was motivated by the expectation that the inflaton decay rate is proportional to its mass, so that an increase in the latter enhances the number of decay channels into fermion-antifermion pairs. Since inflaton decays in the unbroken phase generically violate baryon number, such an enhancement leads to a corresponding amplification of the baryon asymmetry.
Moreover, we adopted a Yukawa-like non-minimal coupling between the scalar curvature and the spectator field, in line with recent findings in the literature showing that such interactions can significantly enhance baryon asymmetry production relative to the purely minimal coupling case.

As in the standard spontaneous baryogenesis, both the inflaton and the standard model fermions were assumed to be minimally coupled to gravity, whereas for the spectator field was allowed a non-minimal interaction of the Yukawa type with $R$. This non-minimal term was expected to provide the dominant contribution to the departure from the background result in the baryon asymmetry. We adopted a semiclassical approach in which all the underlying fields, except the fermions, are essentially treated classically, with the baryonic degrees of freedom effectively included into the fermionic sector only.

The equations of motion for the inflaton and the spectator field were solved perturbatively: to zeroth and first order in the dimensionless coupling parameter $\xi$, respectively. Additionally, we expanded to first order in the Yukawa-like coupling $\sigma$, and, initially, neglected the expansion of the Universe. The inflaton field solution was obtained in the limit $\Omega > m_{\phi}$, keeping in particular terms up to first order in $m_{\phi} / \Omega$. The decay rate was assumed to satisfy $\Gamma \ll \Omega$, in accordance with the assumptions of the background model. The so-obtained solution contains first-order corrections in both $\xi$ and $\xi \sigma$, as expected.

We then computed the corresponding baryon asymmetry generated by the corrected inflaton profile, under the assumption $\Gamma \ll \Omega$, and retaining terms up to first order in $m_{\phi} / \Omega$. The resulting corrective factor to the background asymmetry is
$\{1+ \Xi[1+64\pi^2 \alpha \Sigma/(3g^4)]\}$,
where the first term corresponds to the standard background contribution, the second arises from the spectator-induced correction in absence of the Yukawa-like term, and the third captures the leading effect of the non-minimal curvature coupling.

We showed that \emph{without the Yukawa-like interaction the correction from the spectator is negligible, whereas the inclusion of the non-minimal term leads to a substantial first-order contribution which, depending on the model parameters, may become comparable to or larger than the background value}. However, even enhancing the baryogenesis, considering a scalar spectator field was proved to be not enough in order to faithfully reproduce the experimental value of the baryon asymmetry. This output, found in previous literature, see e.g. Ref.~\cite{DeSimone:2016ofp}, suggests the need of extending the spontaneous baryogenesis paradigm, by adding additional symmetries and including effects beyond the standard model of particle physics.

Future works include further extensions of the spontaneous baryogenesis framework, following the inclusion of electromagnetic interactions through a coupling of the inflaton to the electromagnetic field. Another intriguing possibility is the introduction of a disformal non-minimal coupling between the inflaton and the metric, as a generalization of the better-known conformal coupling $R \theta^2$, already explored in the literature. It would also be worthwhile to replace the complex scalar spectator with a fermionic spectator field to assess its impact on the baryogenesis mechanism. Lastly, a more realistic treatment of sponteneous baryogenesis would be needful by incorporating quantum chromodynamics and may unveil additional mechanisms of asymmetry generation.

\begin{acknowledgments}
OL acknowledges financial support from the Fondazione  ICSC, Spoke 3 Astrophysics and Cosmos Observations. National Recovery and Resilience Plan (Piano Nazionale di Ripresa e Resilienza, PNRR) Project ID CN$\_$00000013 "Italian Research Center on  High-Performance Computing, Big Data and Quantum Computing"  funded by MUR Missione 4 Componente 2 Investimento 1.4: Potenziamento strutture di ricerca e creazione di "campioni nazionali di R$\&$S (M4C2-19 )" - Next Generation EU (NGEU)
GRAB-IT Project, PNRR Cascade Funding
Call, Spoke 3, INAF Italian National Institute for Astrophysics, Project code CN00000013, Project Code (CUP): C53C22000350006, cost center STI442016. OL is also thankful to Antonio Capolupo and  Aniello Quaranta for insightful discussion on the topic of baryogenesis. MM acknowledges support by the project OASIS, “PNRR Bando a Cascata da INAF M4C2 - INV. 1.4”. MM is also in debit to Stefano Mancini
\end{acknowledgments}

\appendix

\section{\label{app1}Dirac theory in FRW spacetime}

The free Dirac equation in curved spacetime is
\begin{equation}
i\gamma^{\mu}\nabla_{\mu}\psi-m\psi= ie^{\mu}_a\gamma^{a}(\partial_{\mu}+\Omega_{\mu})\psi-m\psi=0,
\label{eqAPP1}
\end{equation}
that in a FRW universe is simplified to
\begin{align}
\nonumber
& i\gamma^0\dot{\psi}+\frac{i}{a}\gamma^i\partial_i\psi+i\gamma^i\frac{\dot{a}\gamma_i\gamma_0}{2a}=m\psi\implies\\
& i\gamma^0\bigg{(}\dot{\psi}+\frac{3H}{2}\psi\bigg{)}+i\gamma^i\frac{\partial_i\psi}{a}=m\psi.
\label{eqAPP2}
\end{align}

It is convenient to define $\tilde{\psi}$ such that $\psi=a^{3/2}\tilde{\psi}$. For the new field $\tilde{\psi}$, we obtain
\begin{align}
\nonumber
&i\gamma^0\bigg{(}-\frac{3}{2}a^{-3/2}H\tilde{\psi}+a^{-3/2}\dot{\tilde{\psi}}+\frac{3}{2}Ha^{-3/2}\tilde{\psi}\bigg{)}+\\
\nonumber
&+a^{-3/2}i\gamma^i\frac{\partial_i\tilde{\psi}}{a}=ma^{-3/2}\tilde{\psi}\implies\\
&i\gamma^0\dot{\tilde{\psi}}+i\gamma^i\frac{\partial_i\tilde{\psi}}{a}=m\tilde{\psi}.
\label{eqAPP3}
\end{align}

Eq. (\ref{eqAPP3}) is the free Dirac equation with the spatial components of the 4-momentum rescaled by a factor $1/a$. In the regime of negligible expansion of the universe\footnote{In this phase, we are neglecting the expansion of the universe.}, $a$ can be consider just a rescaling parameter. Therefore, the canonically quantized solution of Eq. (\ref{eqAPP3}) is the same as in Minkowski spacetime but with the momentum $\vec{p}$ rescaled by a factor $1/a$, that is
\begin{widetext}
\begin{equation}
\tilde{\psi}(t,\vec{x})=\sum_s\int\frac{d^3p'}{(2\pi)^3}\frac{1}{\sqrt{2E'_p}}\bigg{(}u(\vec{p'},s)a(\vec{p'},s)e^{-i(E'_pt-\vec{p'}\cdot \vec{x})}+v(\vec{p'},s)b^{\dag}(\vec{p'},s)e^{i(E'_pt-\vec{p'}\cdot\vec{x})}\bigg{)},
\label{eqAPP4}
\end{equation}
\end{widetext}
where $\vec{p'}=\vec{p}/a$ and $E'^2_p=p'^2+m^2=p^2/a^2+m^2$. By a rescaling of the 3D-momentum $\vec{p'}\to\vec{p}=a\vec{p'}$, we get back for $\tilde{\psi}$ the canonical quantization of the Dirac field in Minkowski spacetime. This should not be surprising, because summing over all the 3D-momenta makes the Dirac field be independent of any rescaling of $\vec{p}$. For $\psi$, thus, we obtain
\begin{widetext}
\begin{equation}
\psi(t,\vec{x})=a^{-3/2}(t)\sum_s\int\frac{d^3p}{(2\pi)^3}\frac{1}{\sqrt{2E_p}}\bigg{(}u(\vec{p},s)a(\vec{p},s)e^{-i(E_pt-\vec{p}\cdot \vec{x})}+v(\vec{p},s)b^{\dag}(\vec{p},s)e^{i(E_pt-\vec{p}\cdot\vec{x})}\bigg{)}.
\label{eqAPP5}
\end{equation}
\end{widetext}

For the purposes of this work, it is also interesting to report the expression of the stress-energy tensor associated to a fermion in FLRW spacetime,
\begin{widetext}
\begin{equation}
\begin{split}
T_{\mu\nu}^{(\text{f})}=-\frac{i}{2}g_{\mu\nu}\bigg{(}\overline{\psi}\gamma^{\alpha}\nabla_{\alpha}\psi-\nabla_{\alpha}(\overline{\psi}\gamma^{\alpha})\psi\bigg{)}+g_{\mu\nu}m\overline{\psi}\psi+\frac{i}{4}\bigg{(}\overline{\psi}\gamma_{\mu}\nabla_{\nu}\psi+\overline{\psi}\gamma_{\nu}\nabla_{\mu}\psi-\nabla_{\nu}(\overline{\psi}\gamma_{\mu})\psi-\nabla_{\mu}(\overline{\psi}\gamma_{\nu})\psi\bigg{)},
\end{split}
\label{eqAPPenimpu}
\end{equation}
\end{widetext}
whose trace is thus
\begin{equation}
T^{\mu\text{\ \ (f)}}_{\mu}=-\frac{3i}{2}\bigg{(}\overline{\psi}\gamma^{\mu}\nabla_{\mu}\psi-\nabla_{\mu}(\overline{\psi}\gamma^{\mu})\psi\bigg{)}+4m\overline{\psi}\psi.
\label{eqAPPenimptrace}
\end{equation}

In so doing, the full stress-energy tensor, $T_{\mu\nu}$, appearing in Eq. (\ref{eqGRAV1}), results

\begin{widetext}
\begin{equation}
\begin{split}
T_{\mu\nu}&=-\frac{f^2}{2}g_{\mu\nu}(\nabla^{\alpha}\theta)(\nabla_{\alpha}\theta)+g_{\mu\nu}V(\theta)+f^2(\nabla_{\mu}\theta)(\nabla_{\nu}\theta)-\frac{i}{2}g_{\mu\nu}\left[\overline{Q}\gamma^{\alpha}\nabla_{\alpha}Q-\nabla_{\alpha}(\overline{Q}\gamma^{\alpha})Q\right]+g_{\mu\nu}m_Q\overline{Q}Q+\\&+\frac{i}{4}\left[\overline{Q}\gamma_{\mu}\nabla_{\nu}Q+\overline{Q}\gamma_{\nu}\nabla_{\mu}Q-\nabla_{\nu}(\overline{Q}\gamma_{\mu})Q-\nabla_{\mu}(\overline{Q}\gamma_{\nu})Q\right]-\frac{i}{2}g_{\mu\nu}\left[\overline{L}\gamma^{\alpha}\nabla_{\alpha}L-\nabla_{\alpha}(\overline{L}\gamma^{\alpha})L\right] + g_{\mu\nu}m_L\overline{L}L+\\&+\frac{i}{4}\left[\overline{L}\gamma_{\mu}\nabla_{\nu}L+\overline{L}\gamma_{\nu}\nabla_{\mu}L-\nabla_{\nu}(\overline{L}\gamma_{\mu})L-\nabla_{\mu}(\overline{L}\gamma_{\nu})L\right]+g_{\mu\nu}(\nabla_{\alpha}\theta)J^{\alpha}-(\nabla_{\mu}\theta)J_{\nu}-(\nabla_{\nu}\theta)J_{\mu}+\\&-g_{\mu\nu}\frac{gf}{\sqrt{2}}(\overline{Q}L+\overline{L}Q) + 2(\nabla_{\mu}\phi)(\nabla_{\nu}\phi)-g_{\mu\nu}\left[(\nabla^{\alpha}\phi)(\nabla_{\alpha}\phi)-m^2_{\phi}\phi^2-\xi f^2\theta^2\phi^2\right],
\end{split}
\label{eqENIMPTENSOR}
\end{equation}
\end{widetext}

whose trace, appearing in Eq. (\ref{eqGRAV2}), is thus
\begin{widetext}
\begin{equation}
\begin{split}
T^{\mu}_{\mu}&=-f^2(\nabla^{\alpha}\theta)(\nabla_{\alpha}\theta)+4V(\theta)-\frac{3i}{2}\left[\overline{Q}\gamma^{\mu}\nabla_{\mu}Q-\nabla_{\mu}(\overline{Q}\gamma^{\mu})Q\right] + 4m_Q\overline{Q}Q-\frac{3i}{2}\left[\overline{L}\gamma^{\mu}\nabla_{\mu}L-\nabla_{\mu}(\overline{L}\gamma^{\mu})L\right] +\\&+4m_L\overline{L}L+2(\nabla_{\alpha}\theta)J^{\alpha}-4\frac{gf}{\sqrt{2}}(\overline{Q}L+\overline{L}Q)-2(\nabla^{\alpha}\phi)(\nabla_{\alpha}\phi)+4m^2_{\phi}\phi^2+4\xi f^2\theta^2\phi^2.
\end{split}
\label{eqENIMPTENSORTRACE}
\end{equation}
\end{widetext}

\section{\label{app3}The produced baryon asymmetry}

In order to compute the baryon asymmetry, we have to compute the integral of $\omega^2$ multiplied by twice the real part of the expression in Eq. (\ref{eqPRODUCTTHETATILDE}) divided by the imaginary unity, with $\omega$ as integration variable,
\begin{widetext}
\begin{equation}
\int_0^{+\infty}\omega^2\cdot2\text{Re}\bigg{\{}\frac{\tilde{\theta}(2\omega)\tilde{\theta}^{2*}(2\omega)}{i}\bigg{\}}d\omega=
\Omega^3\int_0^{+\infty}x^2\cdot2\text{Re}\bigg{\{}\frac{\tilde{\theta}(2\omega)\tilde{\theta}^{2*}(2\omega)}{i}\bigg{\}}dx,
\label{eqNEWAPP1}
\end{equation}
\end{widetext}
where we have to take in mind that $\omega=x\Omega$. By using the extended form of $\tilde{\theta}(2\omega)\tilde{\theta}^{2*}(2\omega)$ in Eq. (\ref{eqPRODUCTTHETATILDE}) and by virtue of Eq. (\ref{eq38}), we get

\begin{widetext}
\begin{equation}
\begin{split}
&n_B=\frac{2g^2f^2\Omega^3}{2\pi^2}\int_0^{+\infty}x^2\text{Re}\bigg{\{}\frac{\tilde{\theta}_0(2\omega)\tilde{\theta}_0^{2*}(2\omega)}{i}+\Xi\bigg{[}\frac{\tilde{\theta}_0^{2*}(2\omega)\tilde{\theta}_1(2\omega)+2\tilde{\theta}_0(2\omega)\tilde{\theta}_{01}^{2*}(2\omega)}{i}-\alpha\frac{\tilde{\theta}_0^{2*}(2\omega)\tilde{\theta}_2(2\omega)+2\tilde{\theta}_0(2\omega)\tilde{\theta}_{02}^{2*}(2\omega)}{i}\bigg{]}+\\&+\Xi\Sigma\bigg{[}\frac{\tilde{\theta}_0^{2*}(2\omega)\tilde{\theta}_3(2\omega)+2\tilde{\theta}_0(2\omega)\tilde{\theta}_{03}^{2*}(2\omega)}{i}-\alpha\frac{\tilde{\theta}_0^{2*}(2\omega)\tilde{\theta}_4(2\omega)+2\tilde{\theta}_0(2\omega)\tilde{\theta}_{04}^{2*}(2\omega)}{i}\bigg{]}\bigg{\}}dx=\\&=n_B^{(0)}+\frac{g^2f^2\Omega^3\Xi}{\pi^2}\int_0^{+\infty}x^2\text{Re}\bigg{\{}\bigg{[}\frac{\tilde{\theta}_0^{2*}(2\omega)\tilde{\theta}_1(2\omega)+2\tilde{\theta}_0(2\omega)\tilde{\theta}_{01}^{2*}(2\omega)}{i}-\alpha\frac{\tilde{\theta}_0^{2*}(2\omega)\tilde{\theta}_2(2\omega)+2\tilde{\theta}_0(2\omega)\tilde{\theta}_{02}^{2*}(2\omega)}{i}\bigg{]}+\\&+\Sigma\bigg{[}\frac{\tilde{\theta}_0^{2*}(2\omega)\tilde{\theta}_3(2\omega)+2\tilde{\theta}_0(2\omega)\tilde{\theta}_{03}^{2*}(2\omega)}{i}-\alpha\frac{\tilde{\theta}_0^{2*}(2\omega)\tilde{\theta}_4(2\omega)+2\tilde{\theta}_0(2\omega)\tilde{\theta}_{04}^{2*}(2\omega)}{i}\bigg{]}\bigg{\}}dx,
\end{split}
\label{eqNEWAPP2}
\end{equation}
\end{widetext}

where $n_B^{(0)}$ recovers the background model and is of course given by the contribution to the integral of $\tilde{\theta}_0(2\omega)\tilde{\theta}^{2*}_0(2\omega)$. By solving the integral, we find
\begin{widetext}
\begin{subequations}
\begin{align}
&\int_0^{+\infty}x^2\text{Re}\bigg{\{}\frac{\tilde{\theta}_0^{2*}(2\omega)\tilde{\theta}_1(2\omega)+2\tilde{\theta}_0(2\omega)\tilde{\theta}_{01}^{2*}(2\omega)}{i}\bigg{\}}dx=\frac{\theta_I^3\pi}{16\Omega^2},
\label{eNEWAPP3}\\
&\int_0^{+\infty}x^2\text{Re}\bigg{\{}\frac{\tilde{\theta}_0^{2*}(2\omega)\tilde{\theta}_2(2\omega)+2\tilde{\theta}_0(2\omega)\tilde{\theta}_{02}^{2*}(2\omega)}{i}\bigg{\}}dx=-\frac{\theta_I^3\alpha\pi}{16\Omega^2},
\label{eqNEWAPP4}\\
&\int_0^{+\infty}x^2\text{Re}\bigg{\{}\frac{\tilde{\theta}_0^{2*}(2\omega)\tilde{\theta}_3(2\omega)+2\tilde{\theta}_0(2\omega)\tilde{\theta}_{03}^{2*}(2\omega)}{i}\bigg{\}}dx=-\frac{\theta_I^3\alpha}{8\Omega^2}\bigg{[}\frac{8}{9}\ln(4)+\frac{5}{6}+\ln\bigg{(}\sqrt{\frac{\gamma^2+4\alpha^2}{\gamma^2+4}}\bigg{)}\bigg{]},
\label{eqNEWAPP5}\\
&\int_0^{+\infty}x^2\text{Re}\bigg{\{}\frac{\tilde{\theta}_0^{2*}(2\omega)\tilde{\theta}_4(2\omega)+2\tilde{\theta}_0(2\omega)\tilde{\theta}_{04}^{2*}(2\omega)}{i}\bigg{\}}dx=-\frac{\theta_I^3\pi}{48\gamma\Omega^2}.
\label{eqNEWAPP6}
\end{align}
\end{subequations}
\end{widetext}

The first two of the above-mentioned integrals are related to the terms just proportional to $\Xi$, whereas the other two are due to the Yukawa-like coupling. In particular, among the first two, the dominant contribution is the first one, because the second one is proportional to $\alpha$ and must also be multiplied by $\alpha$ in Eq. (\ref{eqNEWAPP2}), thus it is of the order $\alpha^2$. Quite the opposite, the contributions due to the Yukawa-like coupling are both of the order of $\alpha$, because one already contains $\alpha$ and the other one must be multiplied by $\alpha$ in Eq. (\ref{eqNEWAPP2}). However, among the two, the last one scales as $1/\gamma$, whereas the other is of the order of the unity. Since $\gamma\ll1$, the last integral provides the biggest contribution to the baryon asymmetry. Thus, finally, among the four integrals above-mentioned, the first one and the third one furnish the dominant contributions. The baryon asymmetry thus results

\begin{widetext}
\begin{equation}
\begin{split}
&n_B=n_B^{(0)}+\frac{g^2f^2\Omega^3\Xi}{\pi^2}\bigg{[}\frac{\theta_I^3\pi}{16\Omega^2}+\alpha\Sigma\frac{\theta_I^3\pi}{48\gamma\Omega^2}\bigg{]}=n_B^{(0)}+\frac{g^2f^2\theta_I^3\Omega\Xi}{16\pi}\bigg{(}1+\frac{\alpha\Sigma}{3\gamma}\bigg{)}=\\&=n_B^{(0)}+n_B^{(0)}\Xi\bigg{(}1+\frac{\alpha\Sigma\Omega}{3\Gamma}\bigg{)}=n_B^{(0)}\bigg{[}1+\Xi\bigg{(}1+\frac{8\pi\alpha\Sigma}{3g^2}\bigg{)}\bigg{]}.
\end{split}
\label{eqNEWAPP7}
\end{equation}
\end{widetext}

\end{document}